\documentclass[12pt]{iopart}
\usepackage{epsfig,graphicx}
\usepackage{epsfig,graphicx}
\usepackage{color}
\usepackage{amssymb}
\usepackage{amstext}
\usepackage{amsthm}
\usepackage{amsfonts}
\usepackage{latexsym}
\usepackage{cancel}
\usepackage{epstopdf}
\usepackage{array}
\usepackage{xfrac}

\def\beq{\begin{equation}}{\it}
\def\eeq{\end{equation}}
\def\beqa{\begin{eqnarray}}{\it}
\def\eeqa{\end{eqnarray}}
 
\def\la{\lambda}

\begin{document}

\title{Robustness of discrete semifluxons in closed Bose-Hubbard chains}

\author{A. Gallem\'{\i}$^{1,2}$, M. Guilleumas$^{1,2}$, 
J. Martorell$^1$, R. Mayol$^{1,2}$, A. Polls$^{1,4}$ and B. Juli\'a-D\'{\i}az$^{1,3,4}$}
\address{$^1$ Departament de F\'isica Qu\`antica i Astrof\'isica,\\
Facultat de F\'{\i}sica, Universitat de Barcelona, E--08028 Barcelona, Spain}
\address{$^2$ Institut de Nanoci\`encia i Nanotecnologia de la Universitat 
de Barcelona (IN$\,^2$UB), E--08028 Barcelona, Spain}
\address{$^3$ ICFO-Institut de Ci\`encies Fot\`oniques, Parc Mediterrani 
de la Tecnologia, E--08860 Barcelona, Spain}
\address{$^4$ Institut de Ci\`encies del Cosmos, 
Universitat de Barcelona, IEEC-UB, Mart\'i i Franqu\`es 1, E--08028
Barcelona, Spain}

\begin{abstract}
We present the properties of the ground state and low-energy excitations 
of Bose-Hubbard chains with a geometry that varies from open to closed 
and with a tunable twisted link. In the vicinity of the symmetric 
$\pi-$flux case the system behaves as an interacting gas of discrete 
semifluxons for finite chains and interactions in the Josephson regime. 
The energy spectrum of the system is studied by direct diagonalization 
and by solving the corresponding Bogoliubov--de Gennes equations. 
The atom-atom interactions are found to enhance the presence of 
strongly correlated macroscopic superpositions of semifluxons. 
\end{abstract}

\pacs{03.75.Lm, 03.75.Mn, 67.85.-d}

\maketitle

\section{\bf Introduction}

One-dimensional chains have attracted a great deal of interest in 
the past due to their simple analytical treatment both for 
bosons~\cite{Cazalilla2011} and fermions~\cite{Giamarchi2004}. 
The interplay between superfluidity and the effect of interactions 
in a one-dimensional system is particularly involved with some 
notable phenomena depending strongly on dimensionality, e.g. 
Tonks-Girardeau physics~\cite{Paredes2004,Haller2009}. Superfluid 
properties have been proved in dragged one-dimensional quantum 
fluids in open geometries~\cite{Debusquois2012}, together with 
its breakdown due to interactions~\cite{Astrakharchik2004}. In closed 
periodic geometries, instead, condensate dragging can lead to a 
persistent current, experimentally observed in 
Refs.~\cite{Ramanathan2011,Moulder2012,Wright2013,Eckel2014}. Such persistent 
currents, which have already been experimentally produced in 
superconducting devices~\cite{Mooij1999,vanderWal2000,FornDiaz2010} 
as well as in polariton condensates~\cite{Sanvitto2010}, opened 
promising lines of research for future quantum computers~\cite{Mooij1999}. 
In particular, states with half a quantum circulation or semifluxon 
states, have been detected in polariton spinor Bose-Einstein 
condensates~\cite{Liu2015}, and have also been theoretically 
proposed in superconducting loops with Josephson 
junctions~\cite{Goldobin2002,Walser2008}. 

The physics of persistent currents~\cite{Wright2013,Ryu2007} is intimately linked to phase 
slip events~\cite{Anderson1966,Astafiev2012}, which have been 
widely discussed in (atomic~\cite{Eckel2014,Wright2013}, 
superconductor~\cite{Langer1967,Arutyunov2008} and helium~\cite{Varoquaux2015})
superfluids. They lead to the appearance of objects with topological 
defects (vortices, solitons)~\cite{MunozMateo2015,Gallemi2016} that 
can provoke counterflow superposition~\cite{Gallemi2016,Aghamalyan2015}. 
Persistent currents may also be produced in ultracold atoms trapped in 
a few sites of an optical lattice~\cite{Aghamalyan2015,Paraoanu2003,Kolovsky2006,Roussou2015}, 
such that one can profit from the 
nice properties of ultracold atomic experiments, namely the isolation 
from the environment and the control over the interactions and 
geometry~\cite{Bloch2008,Lewenstein2012}. In this way we have 
previously shown how topological defects can be produced by manipulating 
the phase-dependence of a single link~\cite{Gallemi2015}. There, 
a key ingredient was added to the Bose-Hubbard trimer: a single tunable 
tunnelling link between two modes. This tunable hopping rate can be eventually 
turned to negative, which in the symmetric configuration induces 
a $\pi$ flux through the trimer, thus producing a two-fold degeneracy 
in the ground state of the single-particle spectrum. These two-fold 
degenerate states can be interpreted as discrete semifluxon states 
and provide the basis for the description of the ground state of 
the system, which is, under certain conditions, a cat state of 
semifluxon-antisemifluxon states. This symmetric $\pi-$flux case is 
gauge equivalent to a rotating trimer with a $\pi/3$ phase change 
between all sites, a configuration which has received attention in 
the past~\cite{Hallwood2006}. It is also gauge equivalent to the melting 
of vortex states studied in a higher part of the spectrum in 
Ref.~\cite{Lee2006}. Our symmetric configuration provides a specific 
gauge, which implies a feasible experimental way of producing 
superpositions of semifluxon states.

In this article, we generalize our earlier studies of three-site 
configurations to general closed Bose-Hubbard chains with any number 
of sites. Forcing one of the links to flip the quantum phase, 
macroscopic superpositions of discrete semifluxon states are found 
to form the degenerate set of ground states of the system for small 
but finite atom-atom interactions. This feature makes this setup 
particularly appealing in contrast to the fully symmetric chains 
considered before~\cite{Paraoanu2003}, where persistent currents 
in closed Bose-Hubbard configurations were studied for excited states. 
There are nowadays a variety of experimental setups capable to 
simulate such chains, e.g. ultracold atomic gases trapped in 
a circular array~\cite{Amico2014}. Beyond ultracold atomic systems, Bose-Hubbard 
Hamiltonians have recently been engineered in experiments with 
coupled non-linear optical resonators~\cite{Eichler2014}, where a 
two-mode Bose-Hubbard dimer has been produced with all relevant 
parameters externally controlled. The extension of these setups 
to three or more modes is a promising open line of research. 
Exciton polaritons provide another experimental root to engineer 
these Hamiltonians, since soon after the two-modes case~\cite{amo13} 
discrete ring optical condensates have also been reported~\cite{Dreismann2014}. 

Our setup demands a good control on the tunnelling rate between two 
sites of a Bose-Hubbard chain, both in strength and phase. Ultracold 
atomic gases have provided several proposals to achieve such dependence 
of the tunnelling terms in the case of external 
modes~\cite{Eckardt2005,Tarruell2012,Szirmai2014}. Another option 
would be to replace the external sites for internal ones, building 
the connected Bose-Hubbard chain of internal atomic sublevels. In this 
case, the real one dimensional system is replaced by an extra dimension 
built from the internal sublevels, in the language of Ref.~\cite{Boada2012}. 
In this case a phase-dependent tunnelling can be obtained through 
the Jaksch-Zoller mechanism~\cite{Jaksch2003}. 

The paper is organized as follows. Section~\ref{sec:bh} introduces 
the Hamiltonian that describes the Bose-Hubbard chain with a tunable 
tunnelling. In Sect.~\ref{sect:single}, we will analyze in detail 
the properties of the single-particle problem. Section~\ref{sect:manybody} 
is devoted to study the coherent ground state of the system and 
excitations over mean-field states. In Sect.~\ref{sect:superposition} 
we present the macroscopic superposition of semifluxon states 
that appears for a given set of parameters, and we investigate 
their robustness in Sect.~\ref{sect:robustness}. We summarize our 
results and provide future perspectives in Sect.~\ref{sect:conclusions}.

\vspace{0.5cm}
\section{Tunable Bose-Hubbard Hamiltonian}
\label{sec:bh}

We consider $N$ ultracold interacting bosons populating $M$ quantum 
states (sites). Following standard procedures~\cite{Lewenstein2012} 
the system is described in the lower-band approximation by the 
Bose-Hubbard (BH) Hamiltonian $\hat{\cal H} =  \hat{\cal T} + \hat{\cal U}$, 
which, under the notation of Ref.~\cite{Paraoanu2003} takes the form:
\beqa
\hat{\cal T} &=& 
- \gamma J \,( \hat{a}^{\dag}(M)\, \hat{a}(1) + {\rm h.c.} ) 
-J \sum_{\la = 1}^{M-1} \left( \hat{a}^{\dag}(\la) \,\hat{a}(\la +1) + {\rm h.c.} \right) \nonumber\\
\hat{\cal U} &=& \frac{U}{2} \sum_{\la=1}^{M} \hat{a}^{\dag}(\la)\, \hat{a}^{\dag}(\la)
\, \hat{a}(\la)\, \hat{a}(\la) \,,
\label{eq:hamiltonian}
\eeqa
where $\hat{a}(\la)$ ($\hat{a}^\dagger(\la)$) are the bosonic annihilation 
(creation) operators for site $\la$, fulfilling canonical commutation 
relations. $J$ is the tunnelling amplitude between consecutive sites and 
the parameter $U$ takes into account the on-site atom-atom interaction,
which is proportional to the $s$-wave scattering length and is assumed
to be repulsive, $U>0$. Attractive atom-atom interactions can also be 
produced and would more naturally lead to macroscopic superposition states 
but they are fragile against instabilities. We will analyse the ground 
state structure and the excitation spectrum as a function of the 
dimensionless parameter $\Lambda \equiv NU/J$, which gives the ratio 
between interaction and tunnelling rate. 

As mentioned in the introduction, a crucial ingredient in our model is the 
presence of a single tunable link. In practice, the tunnelling between sites 
$1$ and $M$ can be varied through a parameter $\gamma$, which is 
taken to be real. This allows to study very different configurations, 
for instance, an open 1D Bose-Hubbard chain when $\gamma=0$, and a 
symmetric non-twisted (twisted) closed chain when $\gamma=1 (\gamma=-1)$. 
In the latter two cases $ \hat{\cal H}$ is particularly simple. It can 
be shown that, for the special cases $\gamma=\pm1$, $\hat{\cal H}$ 
is gauge equivalent to a symmetric Hamiltonian (equal couplings) with 
a total flux $\Phi=0$ and $\Phi=\pi$, respectively: 
\begin{equation}
\hat{\cal H}_{\rm sym}=
-J \sum_{\la=1}^{M} \ \left[ e^{i\Phi/M} \,\hat{a}^\dagger(\la) \, \hat{a}(\la +1) + {\rm h.c.}\right] 
+ \hat{\cal U}  \,.
\label{h2}
\end{equation}
In the previous equation, the site $\la=M+1$ corresponds to site $1$ 
(periodicity of the chain). This many-body Hamiltonian describes a 
Bose-Hubbard 1D circular chain, rotating around its symmetry axis 
with angular frequency $\Phi$~\cite{Hallwood2006,Paraoanu2003,arwas2014}. 
Therefore, the physics of the special $\gamma=\pm1$ cases will be 
essentially equivalent to the ones considered prior. However, it is 
worth emphasizing that such local gauge transformation relating the 
Hamiltonians in Eq.~(\ref{eq:hamiltonian}) and Eq.~(\ref{h2}) does 
not exist for a general value of $\gamma$.

\section{The non-interacting problem}
\label{sect:single}

\subsection{Single-particle problem}

For a single particle the problem reduces to finding the eigenvalues 
$\mu$ and eigenvectors $\chi(\la)$, with $\la =  1,\dots,M$, of $\hat{\cal T}$:
\beq
\left(
\matrix{
\mu & J      & 0   & \cdots & 0 & 0 & \gamma J \cr
     J       & \mu & J      & \cdots & 0   & 0  & 0     \cr
     0       & J   &  \mu   & \cdots & 0  & 0  &0    \cr 
   \vdots & \vdots & \vdots & \ddots & \vdots & \vdots & \vdots \cr
                0 &     0 &    0 & \cdots &  \mu & J  & 0 \cr 
                 0 &     0 &    0 & \cdots &  J & \mu & J \cr 
                    \gamma J & 0 & 0 & \cdots & 0 & J & \mu} 
\right)
\left(
\matrix{
\chi(1) \cr \chi(2) \cr \chi(3)\cr \vdots \cr \chi(M-2) \cr \chi(M-1) \cr \chi(M) 
}
\right)
=0\ .
\label{eq:xse1}
\eeq
Using standard techniques in solving tight binding problems, one finds 
that the solutions can be either even (symmetric): 
$ \chi^{(S)}(\la) = \chi^{(S)}(M+1-\la)$ or odd (antisymmetric): 
$ \chi^{(A)}(\la) = -\chi^{(A)}(M+1-\la)$. They can be written conveniently 
in terms of Bloch phases $\phi$  as
\beqa
\chi^{(S)}(\la) &=& \frac{1}{\sqrt{\cal{N}_+}} 
\  \cos \left[ \left(\frac{M+1}{2}-\la\right) \phi \right]\,, \nonumber \\
\nonumber \\
\chi^{(A)}(\la) &=& \frac{1}{\sqrt{\cal{N}_-}}
\  \sin  \left[ \left(\frac{M+1}{2}-\la\right) \phi \right]  \, ,
\label{eq:xse2}
\eeqa
where the normalization factor of each wave function is 
${\cal N}_\pm=(M \pm \sin (M\phi)/\sin \phi)/2$. The respective Bloch 
phases satisfy the implicit equations, 
\begin{figure}[t]                         
\centering
\includegraphics[width=0.5\columnwidth, angle=-90]{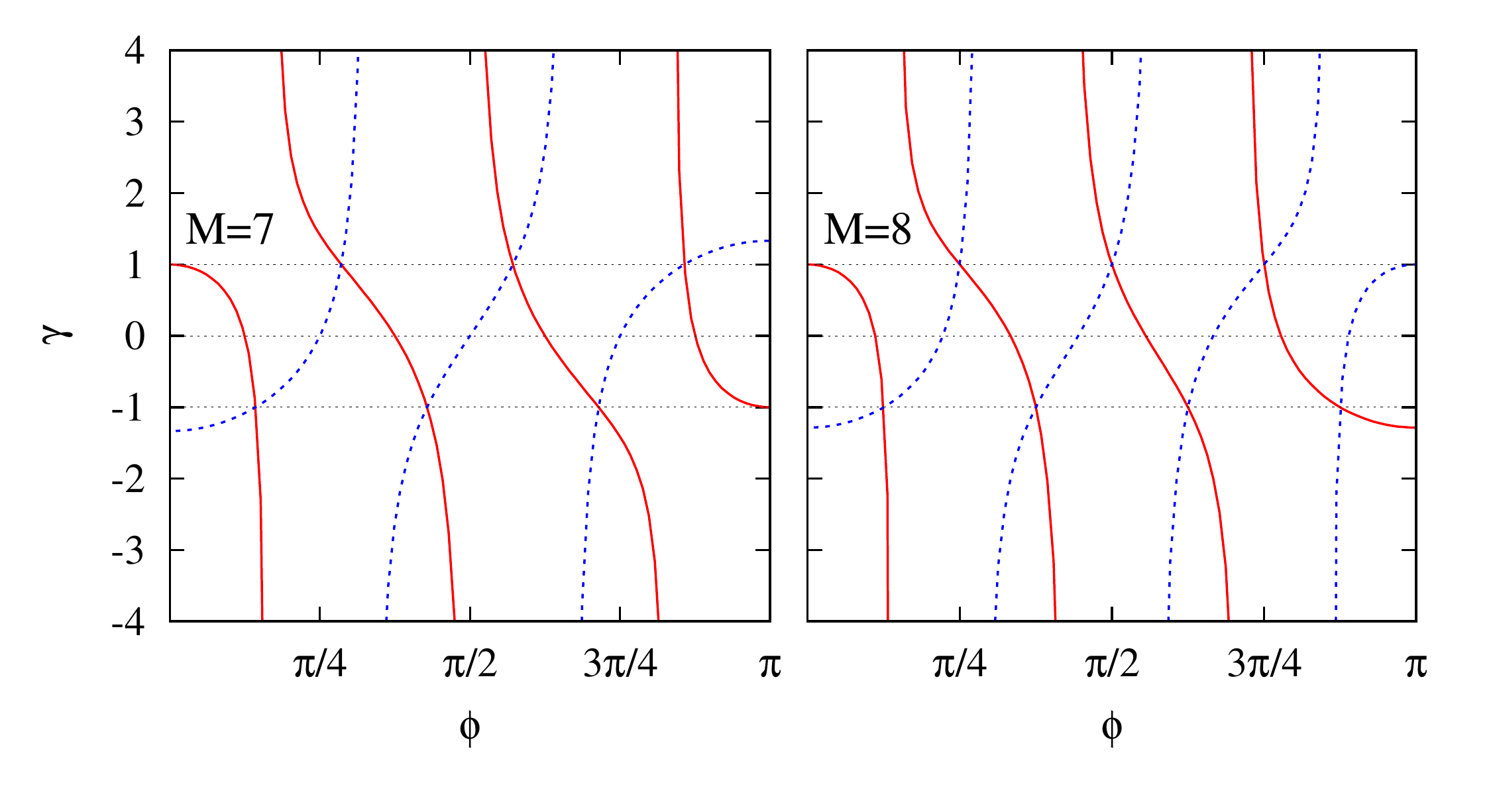}
\caption{Graphical real solutions of Eq.~(\ref{eq:xse3}), even solutions,
solid red lines, and Eq.~(\ref{eq:xse4}), odd solutions, dashed blue 
lines. As en example we depict horizontal lines for $\gamma=\pm1$, dotted 
black lines. For a given $\gamma$, the solutions correspond to the 
crossings between the horizontal dotted line and the solid and dashed 
lines. The number of sites considered is $M=7$ (left) and $M=8$ (right).}
\label{gamm7}
\end{figure} 
\beq
\gamma = \frac{\cos ( \phi(M+1)/2)}{\cos ( \phi(M-1)/2)}
\label{eq:xse3}
\eeq
for the even solutions, and 
\beq
\gamma = - \frac{\sin ( \phi(M+1)/2)}{\sin ( \phi(M-1)/2)}
\label{eq:xse4}
\eeq
for the odd ones. In terms of the Bloch phases, the eigenvalues ($\mu$) 
are then given by
\beq
\mu = -2 J \cos \phi \ .
\label{eq:xse5}
\eeq
Figure~\ref{gamm7} shows the right hand side of the Eqs.~(\ref{eq:xse3}) 
and ~(\ref{eq:xse4}) for symmetric (solid red) and antisymmetric 
(dashed blue) wave functions for a chain with $M=7$ and $M=8$. Drawing a horizontal line at the chosen 
value of $\gamma$, the crossings determine graphically the solutions for real $\phi$, 
as illustrated for the cases $\gamma = 1$ and $-1$. Notice the 
odd/even degeneracies that appear for $\gamma=\pm1$, a characteristic 
feature of these two cases, and their absence for any other values 
of $\gamma$. It can be checked that the $\gamma=+1$ solutions are 
$\phi_q = 2 \pi q /M$, with $q=0,\dots,M-1$, all twice degenerate except the 
first. Similarly, when $\gamma = -1$ the solutions are 
$\phi_q=\frac{2 \pi}{M}(q+\frac{1}{2})$, with  $q=0,\dots,M-1$, all twice degenerate 
except the last.   

\begin{figure}[t]
\centering
\includegraphics[width=0.5\columnwidth,angle=-0]{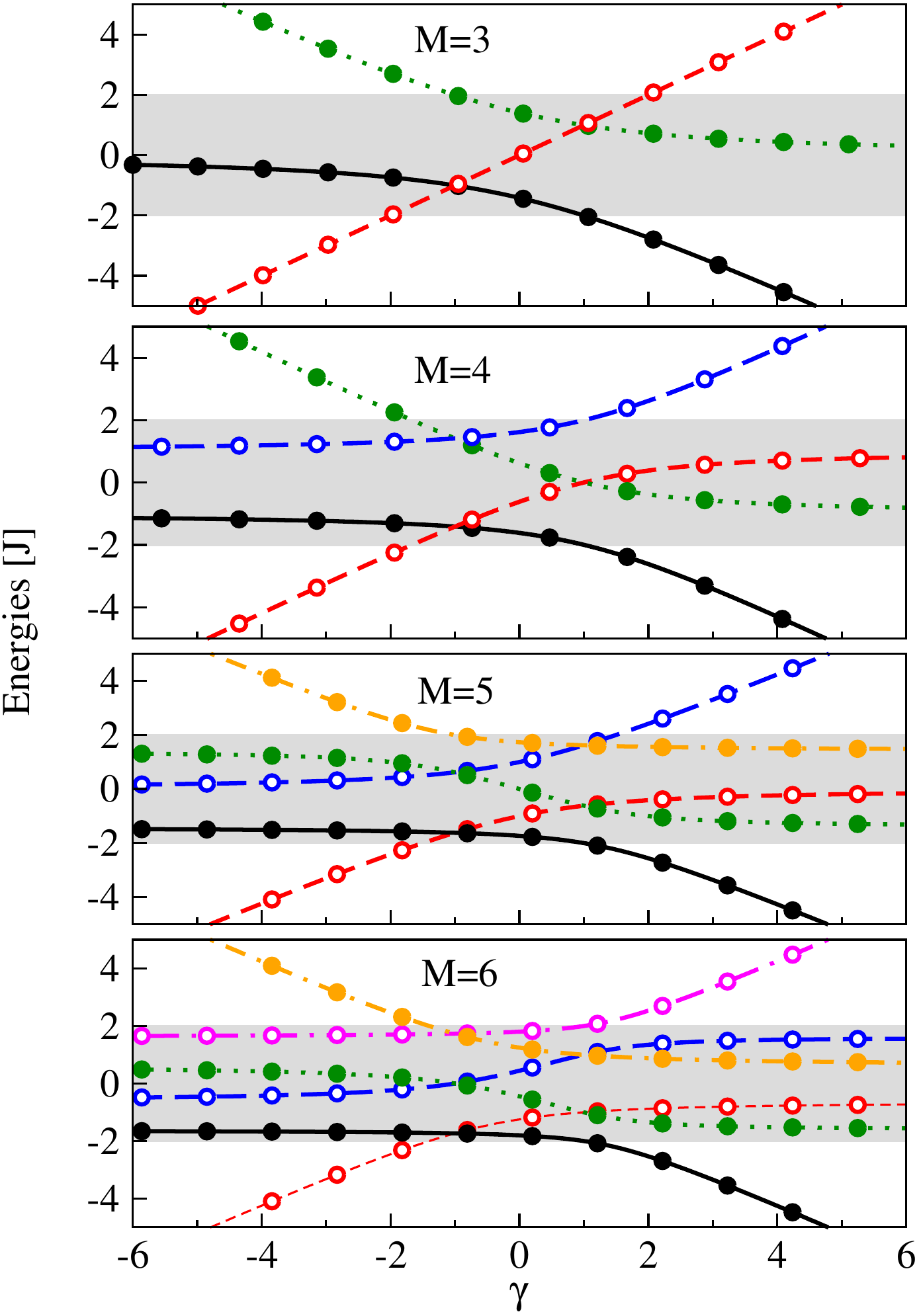}
\caption{Single-particle spectrum for the cases $M=3$, $4$, $5$ and $6$ 
sites, as a function of $\gamma$. Filled (open) symbols characterize levels 
with even (odd)-parity states. The grey region encompasses the range of 
eigenvalues $|\mu|<2J$, which contains all the states with real Bloch 
phase, bulk states. The states outside this region, which have imaginary Bloch phase 
(see the text), are surface states. Energies are in units of $J$. }
\label{specall}
\end{figure}

For arbitrary $\gamma$, all the real solutions that can be extracted from 
those figures fulfill the condition $0 \le \phi \le \pi$. Therefore 
the corresponding energies, Eq.~(\ref{eq:xse5}), are in the band 
$-2J \le \mu \le 2J$, and the associated eigenstates are ``bulk states'', see 
discussion in Sect.~\ref{sect:robustness}. 
This is in agreement with the results in 
Fig.~\ref{specall}, which shows the single-particle spectrum 
as a function of $\gamma$ for $M=3, 4, 5$ and $6$. However, there are 
some special states whose energy is not contained within the previous 
interval. They correspond to ``surface states'', and will be discussed 
later.

The poles of Fig.~\ref{gamm7} (which give the solutions when 
$|\gamma| \to \infty$) are determined by the zeros of the denominators 
in Eqs.~(\ref{eq:xse3}) and~(\ref{eq:xse4}), i.e. 
$\phi^{(S,A)}_{k, (|\gamma| \to \infty)}=\pi k/(M-1)$ where $k$ accounts for the 
odd (even, including $k=0$) natural numbers for the symmetric (antisymmetric) 
wave function. As seen in Fig.~\ref{gamm7} each $\phi$ corresponding to a 
finite $\gamma$ is bounded by those of two consecutive poles 
$\phi_{k-1, (|\gamma| \to \infty)}^{(S,A)}\ < \ \phi^{(S,A)}_k \ < \ \phi_{k, (|\gamma| \to \infty)}^{(S,A)}$,
where $\phi_{-1,(|\gamma| \to \infty)}$ has to be taken as $0$. These inequalities 
show that there are no crossings of single-particle levels of the same 
parity. This feature can be readily seen in Fig.~\ref{specall}, since even 
single-particle levels (filled symbols) do not intersect levels corresponding 
to other even-parity states, independently of the number of sites (and the 
same occurs for odd-parity states, represented by open symbols). Note also 
that the curves associated to the solid red (dashed blue) lines in 
Fig.~\ref{gamm7} are monotonically decreasing (increasing). Thus a monotonic 
variation of $\gamma$ gives also a monotonic variation of its 
corresponding energy in Fig.~\ref{specall}.

\subsection{Extension to imaginary values of $\phi$: Surface states.}

The remaining (up to $M$) eigenstates of Eq.~(\ref{eq:xse1}) correspond to 
surface states and require imaginary values of the 
Bloch phase: $\phi = i \eta$, with $\eta $ real. Then the 
implicit equation for the eigenvalues~(\ref{eq:xse3}), becomes
\beq
\gamma = \frac {\cosh (\eta(M+1)/2)}{\cosh (\eta(M-1)/2)} \,.
\label{eq:eim1}
\eeq
and at small $\eta$, the r.h.s. behaves as $1 +M/2 \  \eta^2$ and leads therefore 
to $\gamma > 1$. For large $\eta$ the r.h.s. rises as $e^{\eta}$ and guarantees 
that there will be solutions for any $\gamma > 1$. 

For the antisymmetric solutions one has to introduce also $\phi=i \eta$ 
and the r.h.s of Eq.~(\ref{eq:xse4}) becomes
\beq
\gamma  = - \frac {\sinh ( \eta(M+1)/2)}{\sinh(\eta(M-1)/2)}
\label{eq:eim2}
\eeq
and for small $\eta$ 
\beq
\gamma \simeq -\frac{M+1}{M-1} \left( 1 + \frac{1}{6} M \eta^2 \right) < -{M+1\over M-1}
\label{eq:eim3}
\eeq
in agreement with the lower bound in Fig.~\ref{gamm7}. The asymptotic behaviour 
for large $\eta$ is now $\gamma \simeq -$ exp$(\eta)$. 

Figure~\ref{gamm7} also shows that there is a similar problem with the 
solutions associated to states with highest $\phi$, corresponding to 
the surface states above the grey region in Fig.~\ref{specall}.
In this case one has to set $\phi = \pi + i \eta$ to find the missing solution. 

In analogy to the expression of $\gamma$ as a function of the imaginary Bloch 
phase, the eigenenergies change as well, $\mu = -2 J \cosh \eta$, which at 
large values of $\eta$ becomes $\mu=\mp 2 J \gamma$, for the even ($-$) and 
odd ($+$) surface states. This explains the asymptotic linear behaviour of 
the energies of these states shown in Fig.~\ref{specall}. 

\subsection{The special $\gamma= \pm1 $ cases}

Fig.~\ref{specall} shows that the cases of $\gamma=\pm1$ exhibit degeneracy 
points where the eigenvectors are linear combinations of the solutions of 
the previous Hamiltonian. Remarkably, these new states are no longer 
currentless. One can construct a basis of flow states (from now on, 
``flow basis''), which can be obtained from the bare states through a unitary 
transformation, 
\beqa
{\tilde \chi}_q(\la; \gamma =+1) &=& \frac{1}{\sqrt{M}} e^{i \frac{2\pi}{M}q \la}  \nonumber \\ 
{\tilde \chi}_q(\la; \gamma =-1) &=& \frac{1}{\sqrt{M}} e^{i \frac{2\pi}{M}(q+1/2) \la}  \ .
\label{eq:es5}
\eeqa
These expressions imply equidistributed particle population within all 
the sites but with a phase variation whose gradient between sites lead 
to an azimutal velocity. 

The $\gamma=1$ case is the commonly considered situation in the literature, 
as it also appears in the usual tight-binding models in condensed matter 
systems. Concerning the currents, there is one important difference between 
$\gamma=1$ and $\gamma=-1$. In the former case, the ground state of the 
system corresponds to $\phi=0$, which is a currentless state. It is always 
non-degenerate and its eigenenergy is $\mu_{gs}=-2 J$, independently of 
the number of sites $M$. The first two excited states are degenerate and 
correspond to $\phi=2\pi/M$ and $-2\pi/M$. They are the discrete version 
of the usual vortex states (also called fluxons) with a circulation of 
$\pm 2 \pi$~\cite{Lee2006,Paraoanu2003}. In contrast, the ground state 
of a BH closed chain with $\gamma=-1$ is always degenerate. It is 
spanned by the eigenvectors corresponding to $q=0 \,(\phi=\pi/M)$ and 
$q=M-1\,(\phi=-\pi/M)$ in Eq.~(\ref{eq:es5}), which are discrete 
semifluxon/antisemifluxon states (half-vortices with $\pm \pi$ circulation, 
see Appendix A for further discussion of their properties). The energy gap 
$\Delta \equiv\mu_{\rm ex}-\mu_{\rm gs}$ between the degenerate ground states and 
the first excitation is 
\begin{equation}
\Delta  =  4J\sin (\pi/M) \, \sin (2\pi/M)\simeq 8 {J \pi^2/ M^2}\,. 
\label{eq:gap}
\end{equation}

As explained above, the $\gamma=-1$ case can be related by a local gauge 
transformation to a BH Hamiltonian in which each hopping induces a 
$\pi/M$ phase, as in Eq.~(\ref{h2}). In that case, considered for instance 
in Ref.~\cite{Hallwood2006} for $M=3$, the degeneracy takes place between 
the fully symmetric ground state for $\gamma=1$, and one of the vortex states.

\section{Many-body coherent states}
\label{sect:manybody}

The eigensolutions of $\hat{\cal T}$ described in the previous Section 
define a new basis for the single particle states, the ``mode'' basis. 
The associated creation and annihilation operators will be written as 
$\hat{b}_q^{\dag}$ and $\hat{b}_q$, $q = 1,\dots,M$, so that
\beqa
\hat{b}_q^{\dag}  &=& \sum_\la \chi_q(\la) \,\hat{a}^{\dag}(\la)\,, \quad \hat{b}_q 
           = \sum_\la \chi^*_q (\la) \,\hat{a}(\la) \ ,
\label{eq:mb1}
\eeqa
with all sums running from $1$ to $M$. From unitarity we have
\beqa
 \sum_\la  \chi_q^*(\la) \chi_{q'} (\la) = 
\delta_{q q'}\,,\quad \sum_q \chi_q^* ( \la') \ \chi_q (\la) = \delta_{\la \la'} \ ,
\label{eq:sd4}
\eeqa
and thus $[\hat{b}_q, \hat{b}^{\dag}_p] = \delta_{qp}$. And also that
$
\hat{a}^{\dag}(\la) = \sum_q \chi^*_q (\la)\ \hat{b}^{\dag}_q$ and 
$
\hat{a}(\la) = \sum_q \ \chi_q( \la )\ \hat{b}_q  \,.
$

The coherent states, defined as
\beqa
|\Psi_q^{(N)}\rangle &\equiv& 
\frac{1}{\sqrt{N!}} (\hat{b}_q^\dag)^N |{\rm vac}\rangle = |0, 0, ..., N,..0\rangle \nonumber \\
\label{eq:sd6}
\eeqa
will be also named mean-field states for $N$ bosons. For the special cases 
when $\gamma = \pm 1$ we will see later that working in the ``flow'' basis, 
Eq.~(\ref{eq:es5}), can be more convenient. We define the corresponding creation 
operators as
\beq
{\tilde b}_q^{\dag}(\gamma=\pm 1) = \sum_{\la} {\tilde \chi}_q(\la;\gamma=\pm1) \ \hat{a}^{\dag}(\la) \ ,
\label{eq:flow1}
\eeq
where the coefficients are given in Eq.~(\ref{eq:es5}). For $\gamma=-1$, the cases 
$q=0$ and $q=M-1$ correspond to the semifluxon, $\hat{b}^{\dag}_{\rm sf}$, and 
antisemifluxon states, $\hat{b}^{\dag}_{\rm asf}$, already discussed.

\subsection{The mean-field ground state}

The expectation value of the Hamiltonian (\ref{eq:hamiltonian}) in a  
coherent state leads to the Gross-Pitaevskii (GP) energy functional, which 
will be useful to obtain the excitations over mean-field coherent states. 
The expectation value of $\hat{\cal T}$ can be readily computed by using
that $\hat a^\dag|n\rangle = \sqrt{n+1}\,|n+1\rangle$ and 
$\hat a|n\rangle = \sqrt{n}\,|n-1\rangle$. Since,  
\beq
 \langle \Psi_q^{(N)}|\hat a^{\dag}(\la +1) \hat a(\la) |\Psi_q^{(N)}\rangle 
= N \chi^*_q (\la+1) \chi_q (\la) \,. 
\label{eq:sd7}
\eeq 
Therefore, the matrix elements of $\hat{\cal{T}}$ are,
\begin{eqnarray}
\langle\Psi_q^{(N)}|{\hat{\cal T}}|\Psi_q^{(N)}\rangle &= &
-JN 
\sum_{\lambda=1}^{M-1} (\chi_q^*(\lambda) \chi_q(\lambda+1) +{\rm c.c.})
\nonumber \\
&&- \gamma J N \left(\chi^*_q(M) \chi_q(1) + \chi^*_q(1) \chi_q(M) \right) \,.
\label{eq:sd9}
\end{eqnarray}

One can analogously compute the expectation value of the interaction part 
of the Hamiltonian to get:
\beq
\langle \Psi_q^{(N)}|\,\hat{\cal U}\,|\Psi_q^{(N)}\rangle = \frac{U}{2} N (N-1) 
\sum_{\la=1}^M |\chi_q(\la)|^4 \,. 
\label{eq:sd14}
\eeq
Once the matrix elements of the Hamiltonian are obtained,  
the corresponding GP mean-field equation can be 
derived by adding a Lagrange multiplier to conserve the norm 
$\mu_q \sum_{\la} |\chi_q(\la)|^2$ and differentiating with respect 
to $\chi^*_q(\la)$. We arrive at the mean-field equations, 
\beqa
\mu_q \chi_q(\la) &=& 
-J (\chi_q(\la -1) + \chi_q(\la+1)) + U\,N\,(N-1)\,|\chi_q(\la)|^2\chi_q(\la)\nonumber \\
\mu_q \chi_q(1) &=& 
-J(\gamma \chi_q(M) + \chi_q(2)) +U\,N\,(N-1)\,|\chi_q(1)|^2\chi_q(1)\,,
\label{eq:sd11}
\eeqa
which in matrix form becomes Eq.~(\ref{eq:xse1}) for the non-interacting 
case. Let us point out that for a consistent derivation one has to assume 
that the $\chi_q(\la)$ in Eq.~(\ref{eq:mb1}) are the ones that follow from 
solving the GP equation including the interaction, $\hat{\cal U}$. In general 
these $\chi_q(\la)$ will be different from the ones found without interaction.

\subsection{Elementary excitations}
\begin{figure} [t]                         
\centering
\includegraphics[width=0.5\columnwidth]{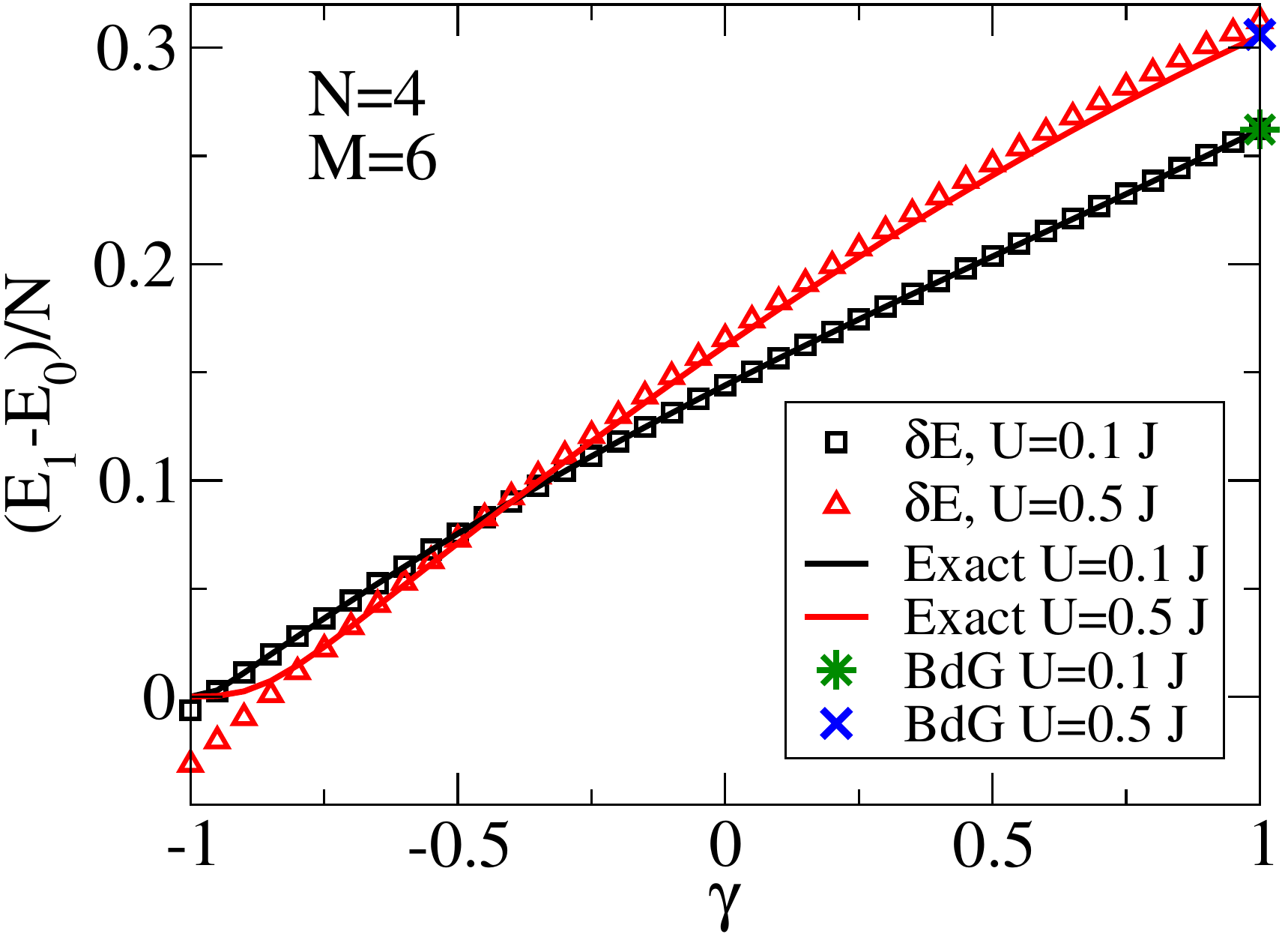}
\caption{Solid lines represent the exact first excitation energy for $U=0.1 J$ (black) 
and $U=0.5 J$ (red). Squares and triangles, first excitation energy computed 
from Eqs.~(\ref{eq:mb6}) and~(\ref{eq:mb9}), replacing $q$ and $p$ by $0$ and 
$1$. Note also that for $-1<\gamma<1$ the ground and first excited states 
change 
symmetry, see Fig.~\ref{gamm7}. The cross and star symbols are the more 
accurate Bogoliubov-de Gennes predictions for $\gamma=1$. In all cases, $N=4$ and $M=6$. }
\label{gpsp}
\end{figure} 

We will now assume that to a good approximation, even when $U \ne 0$, 
the simplest excited states can be approximated as one atom being 
promoted from the coherent state $|\Psi_q^{(N)}\rangle$ made of all the 
atoms occupying the $\chi_q$ to an 
excited orbital $\chi_p$, both $p$ and $q$ orbitals being eigenstates 
of the non-interacting single particle Hamiltonian, 
\beq
|\Psi_q^{(N-1)} \psi_p \rangle \equiv \frac{1}{\sqrt{(N-1)!}} 
(\hat{b}_q^{\dag})^{N-1} \hat{b}_p^{\dag}|\mbox{vac}\rangle \,.
\label{eq:mb4}
\eeq
The expectation value of the Hamiltonian (\ref{eq:hamiltonian}) is the sum of 
the expectation value of the kinetic term and the interactions. The former 
element is: 
\beqa
&&\langle \Psi_q^{(N-1)} \psi_p|{\hat{\cal T}}|\Psi_q^{(N-1)} \psi_p \rangle = 
-J\,\left( \sum_{\la=1}^{M-1} \chi_p^*(\la+1) \chi_p(\la ) 
+ \gamma \chi_p^*(M)\chi_p(1) + {\rm c.c.} \right) \nonumber\\
&&- 
J\,(N-1) \left( \sum_{\la=1}^{M-1} \chi_q^*(\la+1) \chi_q(\la ) 
+ \gamma \chi_q^*(M)\chi_q(1)+ {\rm c.c.} \right)\,.
\label{eq:mb5}
\eeqa
The kinetic energy cost of the promotion of one particle from the $q$ mode 
coherent ground state to the $p$ mode is:
\beqa
 \delta T  &\equiv& \langle \Psi_q^{(N-1)} \psi_p|{\hat{\cal T}}|\Psi_q^{(N-1)} \psi_p\rangle  - 
 \langle \Psi_q^{(N)}|{\hat{\cal T}}|\Psi_q^{(N)} \rangle   \nonumber \\ 
&=& -J  \bigg[ \sum_{\la=1}^{M-1} \left(\chi_p^*(\la+1) \chi_p(\la ) -\chi_q^*(\la+1) \chi_q(\la )\right)
\nonumber\\ 
&&+\gamma (\chi_p^*(M)\chi_p(1)- \chi_q^*(M)\chi_q(1)) + {\rm c.c.}\bigg] \,.
\label{eq:mb6}
\eeqa

A similar procedure can be followed to obtain the expectation value 
of $\hat{\cal U}$ 
\beqa
\langle  \Psi_q^{(N-1)} \psi_p| {\hat{\cal U}}| \Psi_q^{(N-1)} \psi_p\rangle &=&
\frac{U}{2} \bigg[ (N-1)(N-2) \sum_{\la=1}^{M}  |\chi_q(\la)|^4  \nonumber \\ 
&+& 4(N-1) \sum_{\la=1}^M  |\chi_q(\la)|^2 |\chi_p(\la)|^2   \bigg]  
\label{eq:mb8}
\eeqa
%
%
together with the excitation energy cost due to interactions 
$\delta U  \equiv \langle \Psi_q^{(N-1)} \psi_p|{\hat{\cal U}}|\Psi_q^{(N-1)} \psi_p\rangle  - 
 \langle \Psi_q^{(N)}|{\hat{\cal U}}|\Psi_q^{(N)} \rangle$, which yields to 
\beq
\delta U \equiv 
U (N-1)  \sum_{\la=1}^M \bigg[|\chi_q(\la)|^2 ( 2|\chi_p(\la)|^2 -|\chi_q(\la)|^2)  \bigg] \,.
\label{eq:mb9}
\eeq

Figure~\ref{gpsp} shows the calculated lowest excitation energy, 
$\delta E = \delta T + \delta U$ for a range of values of $\gamma$ and 
two particular values of $U$. In 
addition, at $\gamma=1$ the energy gap predicted by the Bogoliubov-de 
Gennes approach computed as in Eqs.~(20) and (21) of~\cite{Paraoanu2003} 
has been added (cross for $U=0.5 \ J$ and star for $U=0.1 \ J$). The 
figure shows that for small values 
of $U$ and except in the vicinity of $\gamma=-1$ the present approximation 
can explain most of the effect of the interaction. The singular point 
$\gamma = -1$ will be discussed in the next Section.

\section{Macroscopic superposition of superconducting flows when $\gamma = -1.$  }
\label{sect:superposition}
\subsection{Two-orbital approximation for the ground state manifold}
\label{ss:two}

As explained above, the non-interacting ground state for the $\gamma=-1$ 
configuration has a two-fold degeneracy between the two semifluxon states. 
In Ref.~\cite{Gallemi2015}, it has been shown that for $M=3$ and small 
interactions, the low-energy states of the system can be described by a 
two-mode model involving only these two single-particle states. For a closed 
chain with $M$ sites, one can expect that the description of the system 
as a macroscopic superposition  of two countercirculating semifluxons 
can be generalized. Moreover, the single-particle energy gap of Eq.~(\ref{eq:gap}) 
can be expected to protect the persistent currents created in the ground 
state manifold in ultracold atomic physics experiments. For $M\lesssim 8$ 
the gap is of the order of the tunnelling $J$. Thus, for few sites and 
small interactions ($N U\lesssim J$), the physics can be restricted to the 
degenerate ground state manifold. 

\begin{figure}[t]
\centering
\includegraphics[width=0.7\columnwidth,angle=-0]{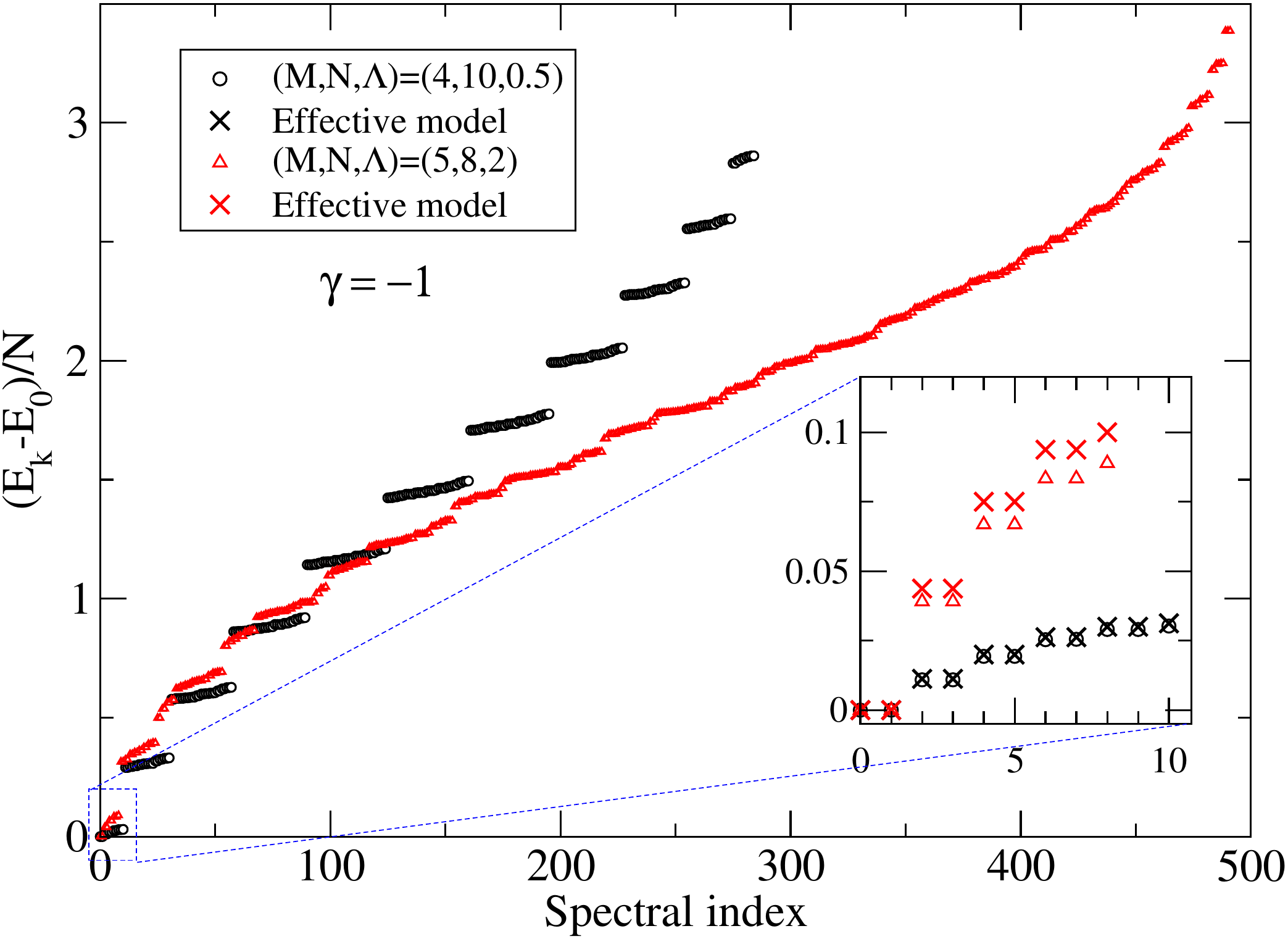}
\caption{Exact many-body energy spectrum (with respect to the ground state 
energy) for $(M,N,\Lambda)=(4,10,0.5)$ (black circles) and $(5,8,2)$ (red 
triangles), in a closed BH chain with $\gamma=-1$. The inset compares the 
exact spectrum with the predicted values using the effective model in 
Eq.~(\ref{eq:effh}) (black and red crosses).}
\label{spec}
\end{figure}

One can generalize the procedure followed in Ref.~\cite{Gallemi2015}, by 
writing the creation and annihilation operators in the coherent flow basis, 
and truncating such decomposition to the semifluxon (sf) and the antisemifluxon 
(asf) states, i.e.,
\beq
\hat a^{\dag}(\la)= \frac{1}{\sqrt{M}} \left( e^{ i \la \pi/M}\, \hat b^{\dag}_{\rm sf} 
                                           + e^{-i \la \pi/M}\, \hat b^{\dag}_{\rm asf} \right) 
\label{eq:tru1}
\eeq 
and the corresponding expression for $\hat a(\la)$. 
One can then rewrite the $\hat{\cal U}$ operator by taking into account 
that sums like $\sum_{\la=1}^M $ exp$(i 2 \la \pi/M)$ vanish, as
\begin{equation}
\hat{\cal U} = 
{U\over 2 M} \left[ \hat{N}^2_{\rm sf} +  \hat{N}^2_{\rm asf} 
+ 4    \hat{N}_{\rm sf} \hat{N}_{\rm asf} 
- (\hat{N}_{\rm sf} + \hat{N}_{\rm asf}) \right]\,,
\label{eq:effh}
\end{equation}
where $\hat{N}_{x}=\hat{b}^\dagger_{x} \hat{b}_{x}$ with $x={\rm sf}$ or 
$x={\rm asf}$ are the number operators of semifluxon or antisemifluxon 
states. Since $N=N_{\rm sf}+N_{\rm asf}$ is constant, the last term in 
Eq.~(\ref{eq:effh}) corresponds to a global energy shift that can be 
neglected. 
The eigenstates of $\hat{\cal U}$ are the new ``Fock''  states
\begin{equation}
|N_{\rm sf},N_{\rm asf} \rangle=
\frac{1}{\sqrt{N_{\rm sf}! N_{\rm asf}!}} 
(\hat{b}^\dagger_{\rm sf})^{N_{\rm sf}} (\hat{b}^\dagger_{\rm asf})^{N_{\rm asf}}  |{\rm vac}\rangle \,.
\label{eq:fluxfock}
\end{equation}

In Fig.~\ref{spec} we plot the energy spectrum obtained by exact diagonalization 
of the many-body Hamiltonian, Eq.~(\ref{eq:hamiltonian}). The band structure of 
the energy spectrum for small interactions, $\Lambda \lesssim 1$, can be 
understood by means of the number of atoms and the degeneracy of the flow basis 
elements (see~\ref{Appendix-A}). In the figure we show two sets of 
parameters $(M,N,\Lambda)$ corresponding to $(4,10,0.5)$ and $(5,8,2)$. The 
spectrum of the first set shows traces of the degeneracy pattern present in the 
non-interacting case whereas for the second set the gaps close in the middle and 
slightly in the upper part of the spectrum. The inset in Fig.~\ref{spec} shows 
the comparison between the low-lying exact many-body spectrum with the prediction 
of Eq.~(\ref{eq:effh}). This  approximation turns out to be very accurate for 
small interactions $\Lambda  < 2$. For the two cases considered in the inset one 
can see that the model provides a good description for $\Lambda = 0.5$, but it 
starts to deviate for larger values, such as $\Lambda=2$.

For small $\Lambda$, the eigenvectors belonging to the low energy manifold can 
be well approximated by
\begin{eqnarray}
|\Psi_{k}^\pm\rangle = {1\over\sqrt{2}} (|k,N-k\rangle \pm |N-k,k\rangle) \,,
\label{cat}
\end{eqnarray}
using the Fock basis, $|N_{\rm sf}, N_{\rm asf}\rangle$. The integer index $k$ runs 
from $0$ to $N$, and the $\pm$ sign labels the two states, which would be degenerate 
in energy if Eq.~(\ref{eq:effh}) was exact. In this approximation, 
$|\Psi_{0}^\pm\rangle$ corresponds to the two-fold degenerate ground state. A 
similar two-orbital approximation and macroscopic superposition of superfluid 
flow was considered in Ref.~\cite{Hallwood2006} for a three well system in a 
different gauge.

\begin{figure}[t]
\centering
\includegraphics[width=0.47\columnwidth,angle=-0]{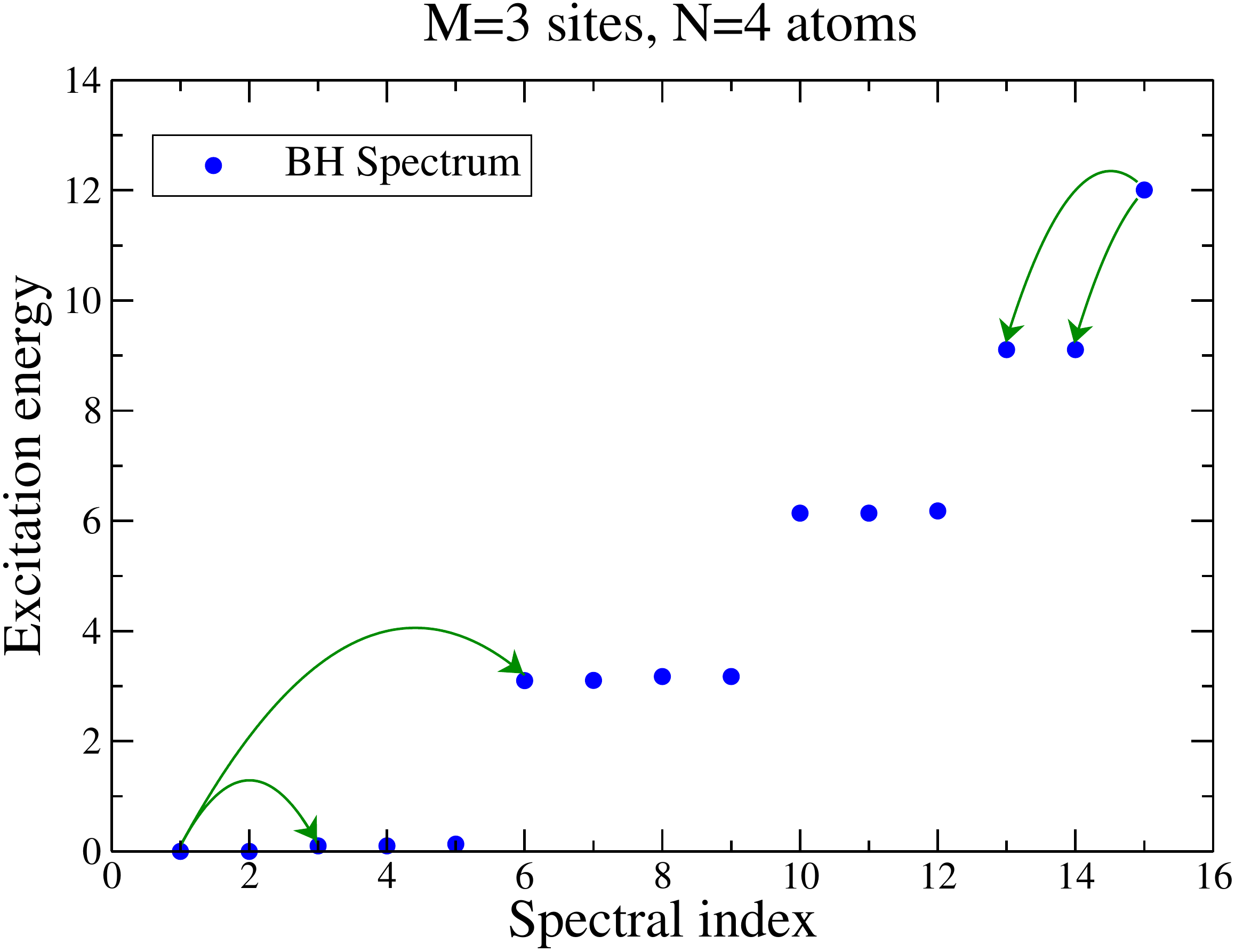}
\includegraphics[width=0.47\columnwidth,angle=-0]{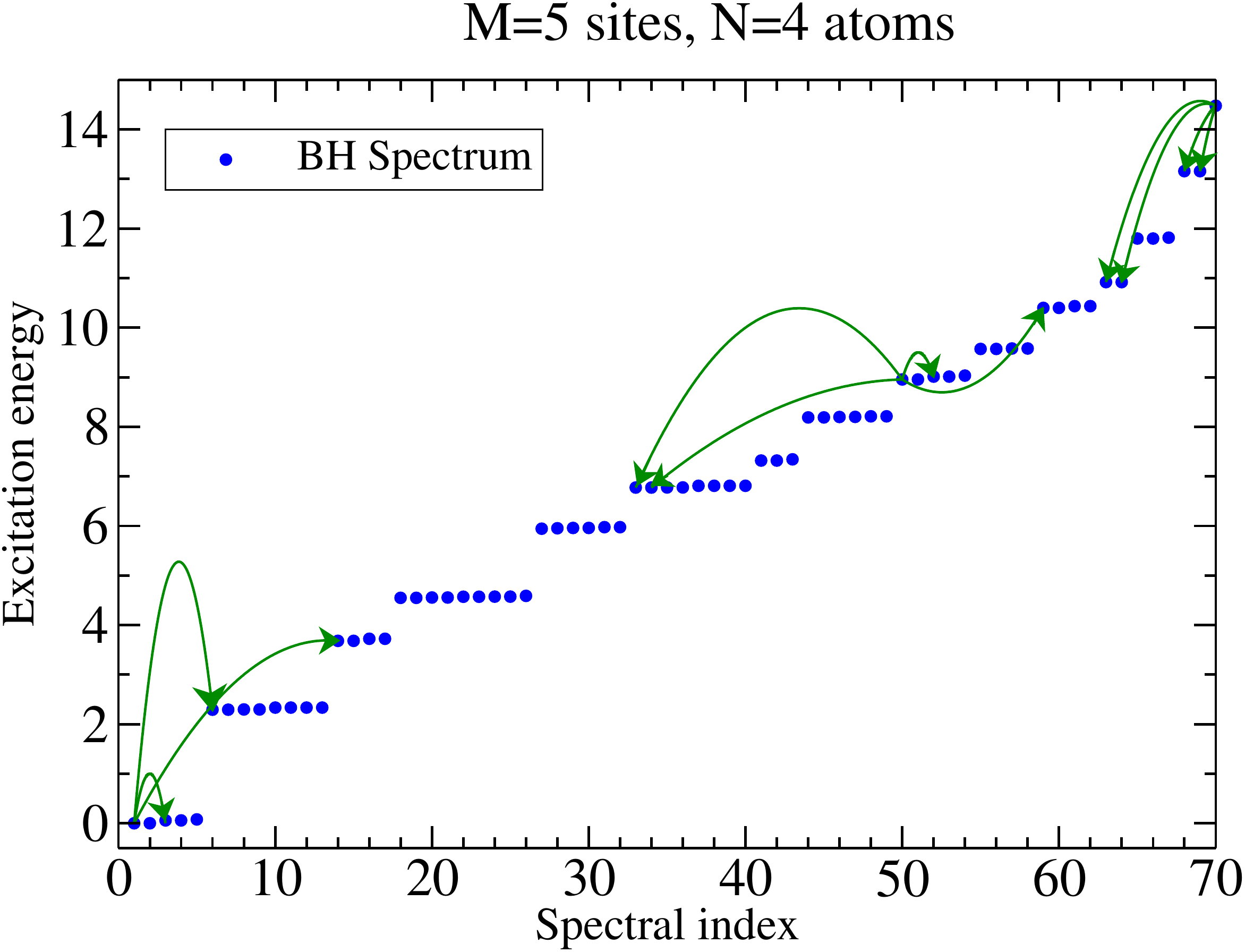}
\caption{The blue dots indicate the exact many-body spectrum 
in a closed BH chain with $\gamma=-1$, for $(M,N,\Lambda)=(3,4,0.4)$ (left), 
and $(5,4,0.4)$ (right). The green arrows, which emerge from coherent 
mean-field states, account for the transitions predicted by the BdG excitations, 
Eq.~(\ref{eq:bogo}). }
\label{fig-M3N4}
\end{figure}

\subsection{Bogoliubov-de Gennes spectrum}

The spectrum of elementary excitations in the weakly interacting regime can 
be studied within the Bogoliubov-de Gennes (BdG) framework. We follow the model 
presented in Ref.~\cite{Paraoanu2003} for a circular array of Bose-Einstein 
condensates with the same tunnelling rate between the sites ($\gamma=1$). 
For the case of $\gamma=-1$, the equations are formally the same but with 
$\phi_q=\frac{2 \pi}{M}(q+\frac{1}{2})$ instead of $\phi_q=\frac{2 \pi}{M}q$
(see the discussion below Eq.~(\ref{eq:xse5})).

The Bogoliubov excitation spectrum is constructed over  mean-field states 
defined as  macroscopically occupied modes in the flow basis, 
Eq.~(\ref{eq:sd6}), with $q=-1,0,1,\dots, M-2$ (or equivalently 
$q=0,1,\dots, M-1$). In the previous 
expression the corresponding coefficients of the creation operator 
$\hat b_q^\dagger$ 
are ${\tilde \chi}_q(\la; \gamma =-1)$ defined in Eq.~(\ref{eq:es5}). 
The periodicity of the system imposes that $q$ is a cyclic index, 
with period $M$: for example, $q=0 \,(-1) $ and $M \,(M-1) $ are equivalent.

The excitation energies relative to a macroscopically occupied state can be 
obtained with the same procedure developed in Ref.~\cite{Paraoanu2003}:
\begin{equation}
E_k^{(\pm)}=
2J\sin\left(\frac{2\pi k}{M}\right) 
\sin(\phi_q)\pm\sqrt{\varepsilon_k \left(\varepsilon_k+\frac{2(N-1)\,U}{M}\right)} \,,
\label{eq:bogo}
\end{equation}
%
with 
\begin{equation}
\varepsilon_k=2J\cos(\phi_q)\left[1-\cos\left(\frac{2\pi}{M}k\right)\right]\,.
\label{bogobogo}
\end{equation}
%
The cyclic index $k$ runs from $0$ to $M-1$, and it is interesting to note that
when $k=0 \,(M)$, the Bogoliubov spectrum yields $E_{k=0}^{(\pm)}=0$, which corresponds 
to remain in the unperturbed mean-field state (without any excitation).

\begin{figure}[t]
\centering
\includegraphics[width=0.8\columnwidth,angle=-0]{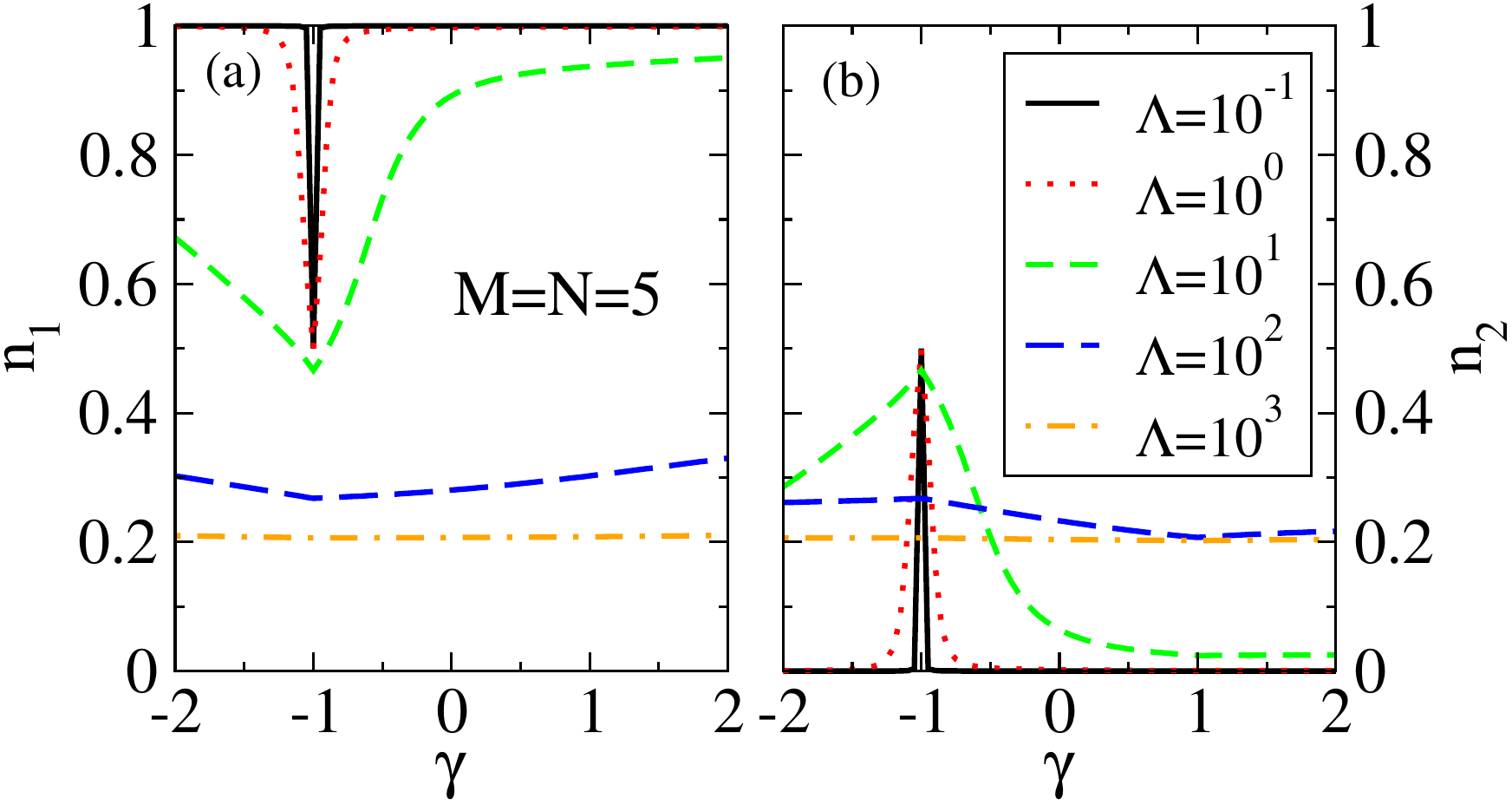}
\caption{(a) 
Largest eigenvalue of the one-body density matrix, the condensed fraction $n_1$. 
(b) Second largest eigenvalue of the one-body density matrix $n_2$. The plots 
correspond to the case of $M=N=5$. The curves have been calculated for values 
of $\Lambda$ that range from $10^{-1}$ to $10^3$, and all of them are represented 
as a function of $\gamma$. }
\label{obdmM5}
\end{figure}

Figure~\ref{fig-M3N4} shows the exact many-body spectrum and the BdG excitations 
(indicated by arrows) relative to the  mean-field like states of a system of 
$N=4$ atoms in $M$ sites. We have calculated the BdG spectrum in a chain with 
$M=3$ and $M=5$ sites with $\gamma=-1$ and for small interactions $U=0.1 \,J$.  
The results are further discussed in~\ref{appb} and collected in 
Tables~\ref{table-1} and~\ref{table-2}.
From the two possible excitations $E_{k}^{(\pm)}$ provided by the BdG calculation 
one has to discard the solution that does not fulfill the BdG normalization 
condition~\cite{Paraoanu2003}  (crossed as \cancel{3.55753} in Tables~\ref{table-1} 
and \ref{table-2}). Moreover, the solutions must fulfill that the 
excitations relative to the ground state are positive, whereas those relative 
to the highly excited macroscopically occupied state in the highest band must be negative.

\begin{figure}[t]
\centering
\includegraphics[width=0.8\columnwidth,angle=-0]{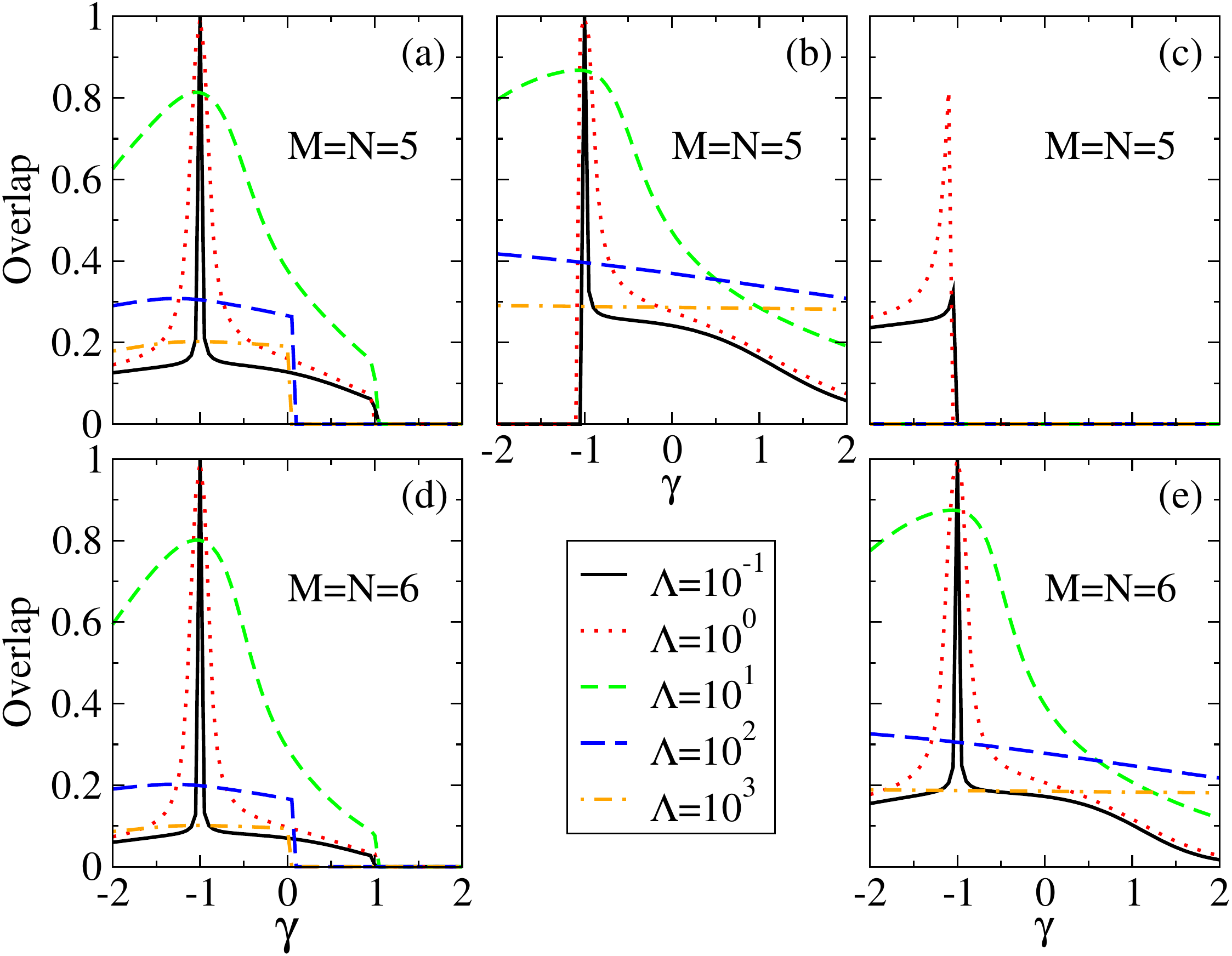}
\caption{Absolute value of the overlap determinant, Eq.~(\ref{eqolap}) for 
$M=N=5$ (a) and $M=N=6$ (d). (b) $|\langle \Psi_{\rm gs}^{(N)} | \Psi_0^+\rangle|$ 
for $M=N=5$. In the case $M=N=6$, 
$|\langle \Psi_{\rm gs}^{(N)} | \Psi_0^+\rangle|\lesssim 10^{-7}$ and is therefore not plotted. 
(c) and (e) depict $|\langle  \Psi_{\rm gs}^{(N)} | \Psi_0^-\rangle|$ for $M=N=5$, and $M=N=6$, 
respectively. The curves have been calculated for values 
of $\Lambda$ that range from $10^{-1}$ to $10^3$, and all of them are represented 
as a function of $\gamma$. }
\label{olapfinal}
\end{figure}

\vspace*{0.5cm}

\section{Robustness of superconducting flows when $\gamma \ne -1$}
\label{sect:robustness}

The macroscopic superpositions of semifluxon states are predicted to appear 
for low interactions in the special $\gamma=-1$ case. Our focus here will be 
on studying the presence of such macroscopic superpositions of superfluid 
flows when $\gamma\neq -1$. Two indicators will be used to signal the presence 
of macroscopic cat-like states as the ones written in Eq.~(\ref{cat}) with 
$k=0$. The first one is the overlap between the cat states and the 
exact solutions resulting from the numerical diagonalization. As 
discussed earlier, for $\Lambda\lesssim 1$ the ground state is almost doubly 
degenerate, then, $\{ |\Psi_0^+\rangle, |\Psi_0^-\rangle\}$ is a suitable basis for 
the lower manifold, as predicted by the two-orbital model.
We therefore define the overlap determinant as, 
\beq
{\cal O} = 
\left|
\left(
\matrix{
 \langle \Psi_{\rm gs}^{(N)}  | \Psi_0^+ \rangle & \langle \Psi_{\rm gs}^{(N)}  | \Psi_0^- \rangle \cr
 \langle \Psi_{\rm 1st}^{(N)}  | \Psi_0^+ \rangle  & \langle \Psi_{\rm 1st}^{(N)}  | \Psi_0^- \rangle  \cr
}
\right) \right|
\,,
\label{eqolap}
\eeq
with $|\Psi_{\rm gs}^{(N)} \rangle$ and  $|\Psi_{\rm 1st}^{(N)} \rangle$ the 
ground and first excited states obtained by direct diagonalization, 
respectively. The overlap ${\cal O}$ is $1$ when the two manifolds are the same.

The second indicator is the fragmentation of the ground state of the system, 
given by the eigenvalues $n_i$ of the single-particle density matrix, 
$\hat{\rho}^{(1)}_{ij}= \langle \Psi| \hat{a}_i^\dagger \hat{a}_j |\Psi\rangle/N$,
with $i,j=1,\dots,M$. The eigenvalues fulfill $\sum_i n_i=1$, and the largest 
one is called the condensed fraction of the system. A fully condensed many-body 
state has $n_1=1$. Instead, the two macroscopic superpositions in Eq.~(\ref{cat}) 
are bifragmented and produce $n_1=n_2=1/2$. A Mott-insulating phase with 
integer filling has $n_i=1/M$, for $i=1,\dots,M$.

\begin{figure*}[t]
\centering
\includegraphics[width=0.4\columnwidth,angle=-90]{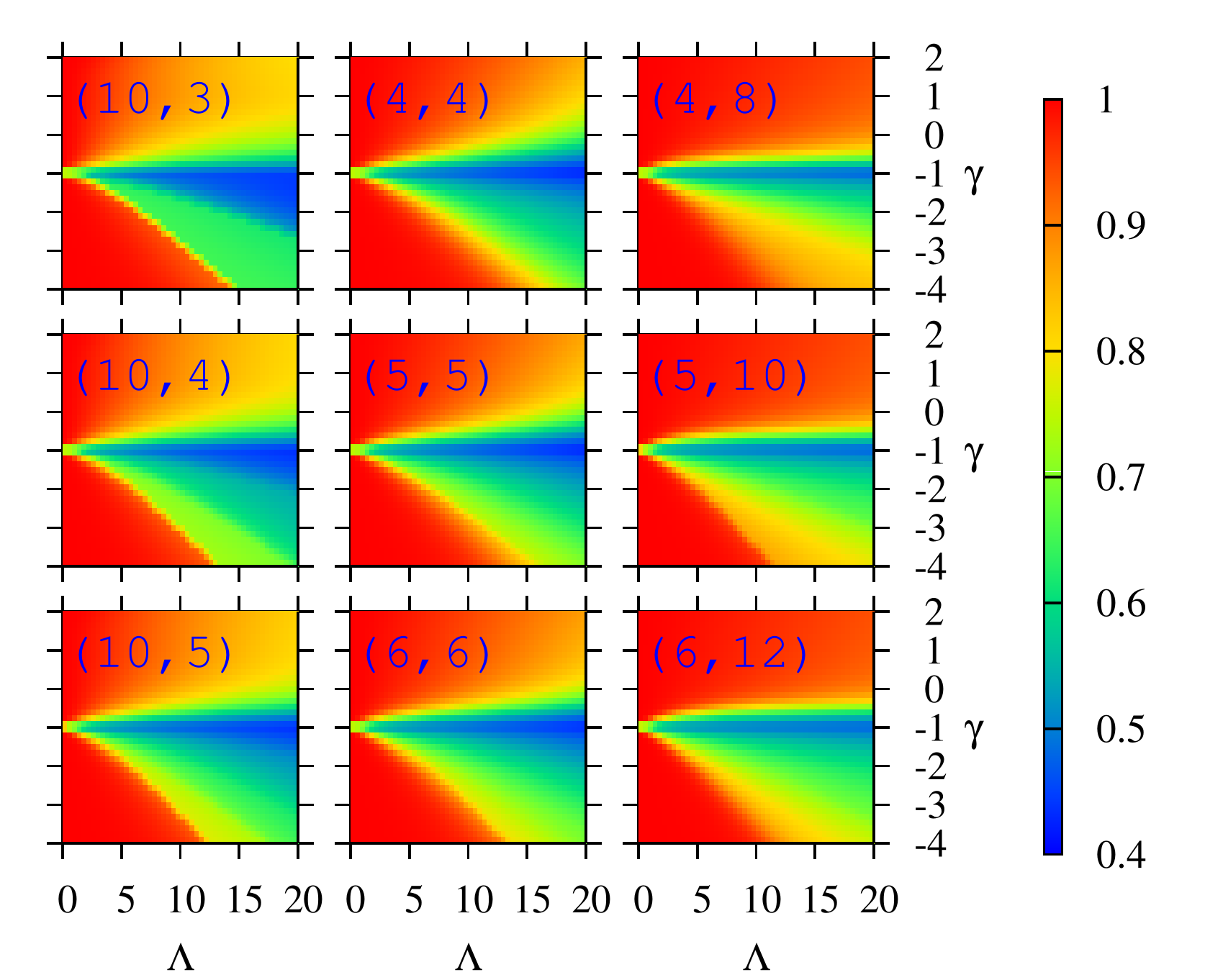}
\includegraphics[width=0.4\columnwidth,angle=-90]{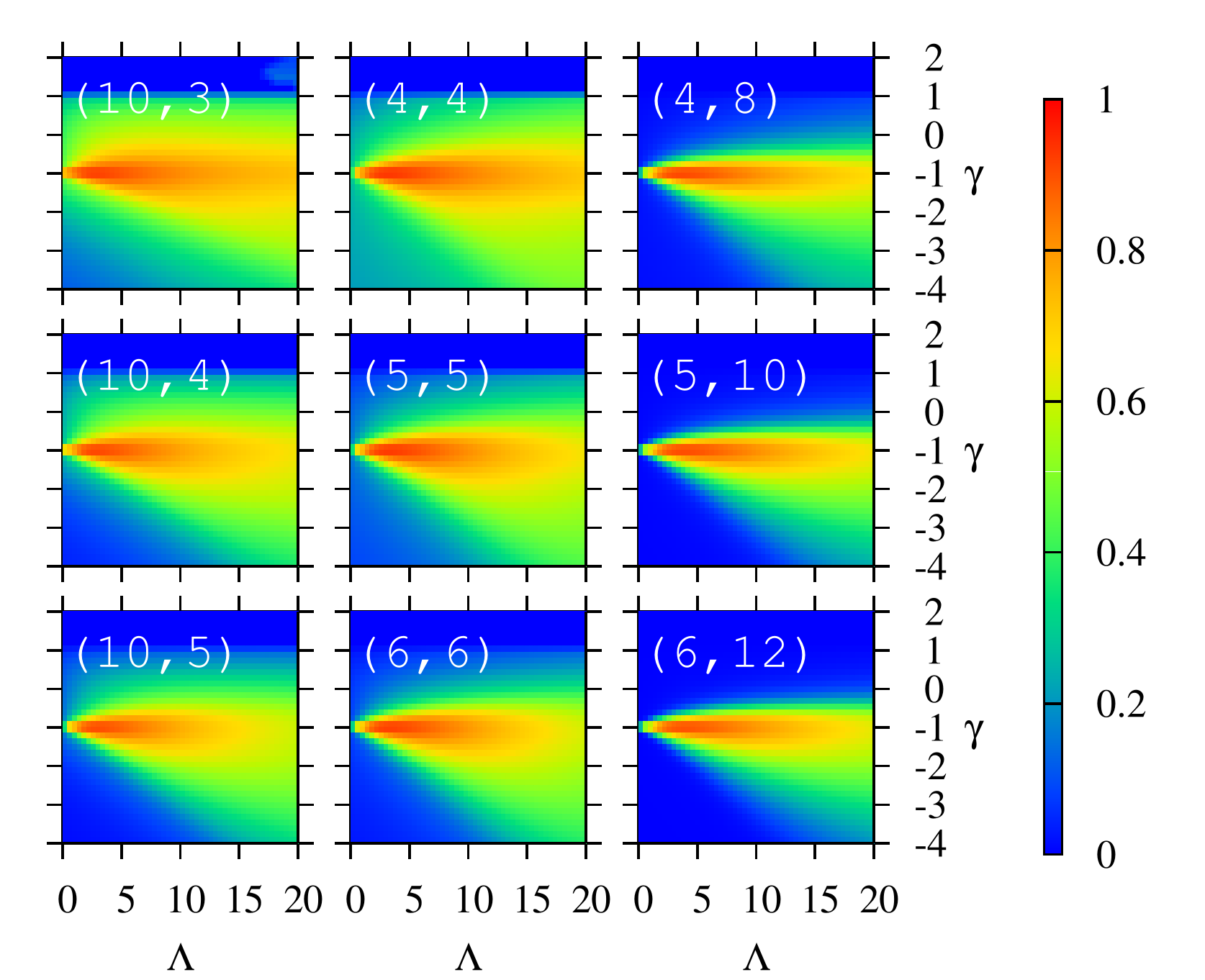}
\caption{The condensed fraction $n_1$ (left panels) and the overlap determinant 
$\mathcal{O}$ (right panels) are represented in a color map as a 
function of $\gamma$ and $\Lambda$, for different values of the pair $(M,N)$, 
which appears as a label in each plot. From top to bottom, the left column has 
increasing filling factor from $3/10$ to $1/2$, and the middle and right columns 
have increasing number of sites $M$ for filling factor $1$ and $2$, respectively.}
\label{map}
\end{figure*}

As expected, the values of the overlap determinant are clearly correlated 
with those of the condensed fraction, see Figs.~\ref{obdmM5},~\ref{olapfinal} 
and~\ref{map}. The system is found to be approximately bifragmented when 
there is a noteworthy overlap determinant. For small $\Lambda$ and $\gamma=-1$ 
we find both condensed fractions $n_1 \simeq n_{2} \simeq 1/2$ and an 
overlap determinant very close to $1$, regardless of the number of particles 
and filling factor, $N/M$. What is somewhat unexpected is that finite 
values of $\Lambda$ enhance the probability of finding these macroscopic 
superposition states when $\gamma\neq -1$. For fixed $M$ and filling 
factors below $1$, the area covered by the significantly fragmented 
configurations broadens for increasing filling, as seen comparing the 
$(M,N)=(10,3)$, $(10,4)$, and $(10,5)$ panels in Fig.~\ref{map}. For 
filling factors larger than one, the fragmented region gets reduced, 
see panels $(M,N=M)$ and $(M,N=2M)$ in Fig.~\ref{map}. 

Note that the overlap determinant would be strictly zero if the quasi 
degeneracy is absent. In this case, even though the ground state may still 
be well described by $|\Psi_0^+\rangle$, we would get a zero overlap between 
the two manifolds. This is the reason why the overlap becomes abruptly 
zero in the vicinity of $\gamma=1$ signalling a level crossing in the many-body 
spectrum between the first and second excited states, see Figs.~\ref{olapfinal} 
and~\ref{map}. This level crossing is directly related to the crossing 
found at $\gamma=1$ in the single-particle spectrum, see Fig.~\ref{specall}. 
Even though the two-fold degeneracy is broken, we find that the actual ground 
state of the system has a sizeable overlap with the $|\Psi_0^+\rangle$ 
state in a broader region of parameters, as seen in 
Fig.~\ref{olapfinal}(b). There, for $N=M=5$ we find that $\Lambda\simeq 1$ is
a good candidate to find counterflow cat states for a broad range of 
$\gamma$. The overlaps between the ground state and the two quasi-degenerate
cat states in the vicinity of $\gamma=-1$ depend critically on the number 
of particles. In the case $N=M=5$ for $\gamma \simeq -1$ we find that the 
overlap between the ground state and $|\Psi_0^+\rangle$  goes 
from almost one for $\gamma\gtrsim -1$ to close to zero for $\gamma\lesssim -1$ 
for $\Lambda\lesssim 1$. While the situation is the opposite for the 
overlap with $|\Psi_0^-\rangle$. A much more symmetric situation is 
found for even number of particles, which does not show this change, 
as seen in Fig.~\ref{olapfinal}(d,e). This behaviour 
can be understood from the two-orbital description of Sect.~\ref{ss:two}, details 
are beyond the scope of the present manuscript. 
\begin{figure}[t]
\centering
\includegraphics[width=0.7\columnwidth,angle=0]{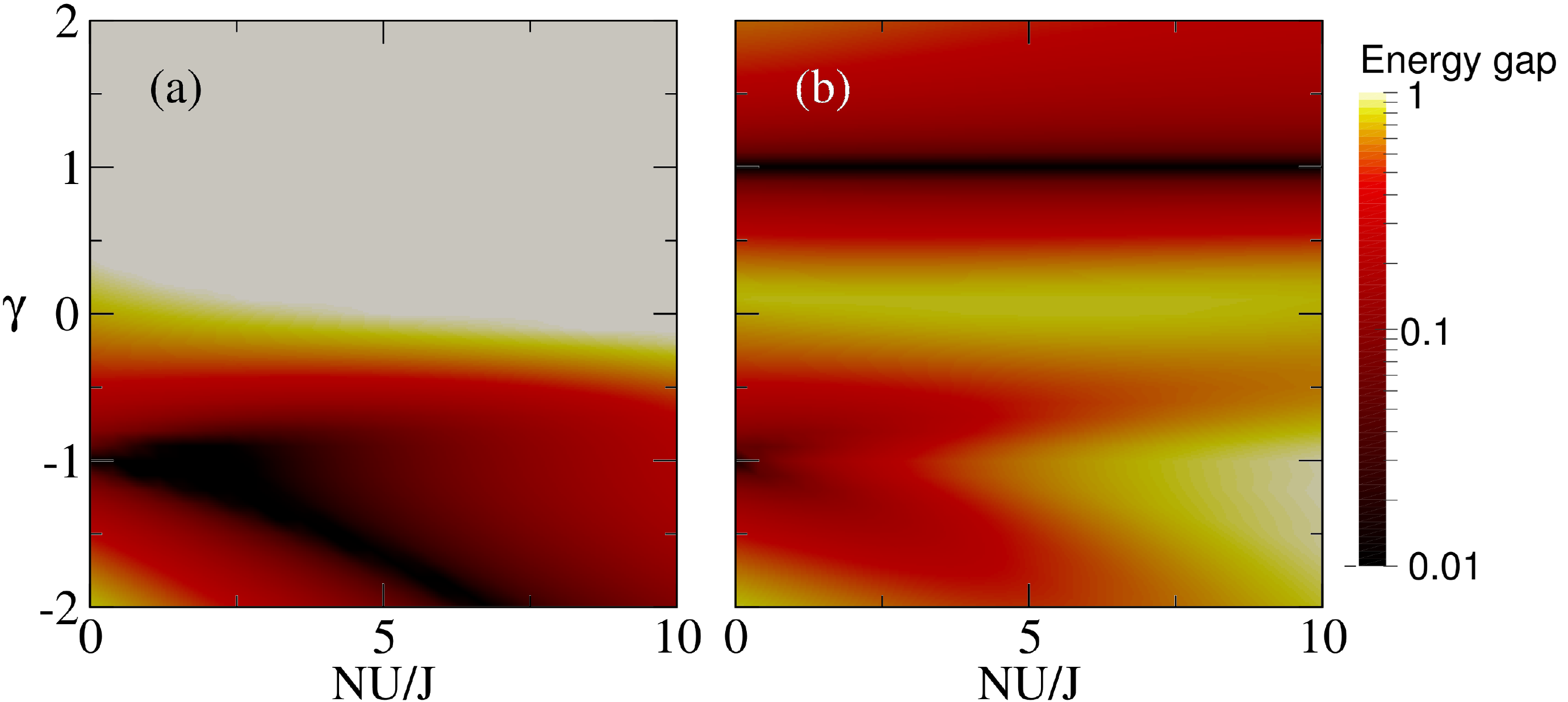}
\caption{(a) Energy gap between the many-body ground state and the first 
excited state, and (b) energy difference between the first and second excited
states. Both are computed as a function of $\Lambda$ and $\gamma$ for $N=M=5$. 
\label{fig:ene}}
\end{figure}

Further increasing the interactions, $\Lambda>100$, the condensed 
fraction decreases quickly to the expected value $1/M$ for the Mott 
insulating regime,  Fig.~\ref{obdmM5}. For integer filling, in the 
$\Lambda \to \infty$ limit the ground state is non-degenerate and 
gapped (Mott insulator of the corresponding filling). Thus, it is 
expected that already for any finite value of $\Lambda$ the two-fold 
degeneracy predicted by the two orbital model would be only approximate. 
For fractional fillings, in contrast, in the large interaction limit 
the ground state will be degenerate, with the surplus particles 
delocalized in the chain. In this case the numerics shows that  the 
two-fold degeneracy predicted by our model is present for all values of 
$\Lambda$.

Besides the condensed fraction and the overlap, a crucial feature 
of a  macroscopic superposition is its robustness, characterized by 
the presence of an energy gap that protects the superposition to be 
affected by excitations that could involve larger-energy excited 
states. In Fig.~\ref{fig:ene} we present the energy difference between 
the ground state and the first excited state, panel (a), and the energy 
difference between the latter and the second excited one, panel (b). There are 
two important properties exhibited by the peculiar case of $\gamma=-1$. 
First, for zero interactions, the ground state has a very large 
degeneracy, $N+1$, stemming from the combinatorial factor, i.e. $N$ 
particles populating two single-particle states in $N+1$ ways, see \ref{Appendix-A}.  For 
very small interactions, $\Lambda\lesssim 0.5$, this is reflected in 
the very small gap $\lesssim 0.01\,J$ both between the ground and 
first, see Fig.~\ref{fig:ene}(a), and between the first and second excited 
many-body eigenstates, see Fig.~\ref{fig:ene}(b). Secondly, for $\gamma=-1$ 
and $\Lambda\gtrsim 5$, a gap starts to open above the 
ground state. This gap opening increases the robustness of the 
macroscopic superposition, which indicates that experimental 
observation of these states should be feasible. 

When $U = 0$ the densities can be easily constructed using 
the wavefunctions given by Eq.~(\ref{eq:xse2}). When $\gamma >-1$, 
the ground state is symmetric, see Fig.~\ref{gamm7}, and the density 
is given by,
\beq
\rho^\gamma(\la) = 
\frac {N}{\cal{N}_+} \ \cos^2 \left[\left(\frac{M+1}{2} - \lambda\right) \phi\right] \,,
\label{eq:den1}
\eeq
with $\phi$  determined from Eq.~(\ref{eq:xse3}). These densities are 
linear in $N$, have a maximum at $\lambda_m = (M+1)/2$ and are site 
symmetric with respect to $\lambda_m$. The simplest example is the 
case $\gamma=0$: then the solution of Eq.~(\ref{eq:xse3}) 
is $\phi=\pi/(M+1) $  and 
\beq
\rho^{\gamma=0}(\la)
= \frac{2N}{M+1} \ \sin^2\left(\frac{\pi}{M+1}\lambda \right) \,,
\label{eq:den2}
\eeq
which shows that at the maximum $\rho_M = 2N/(M+1)$ and at sites $1$ and $M$, 
$\rho= 2N/(M+1) \sin^2 (\pi/(M+1))$. Analogous expressions, involving 
hyperbolic functions, can be easily derived for the surface states.

\begin{figure} [t]
\centering
\includegraphics[width=0.36\linewidth]{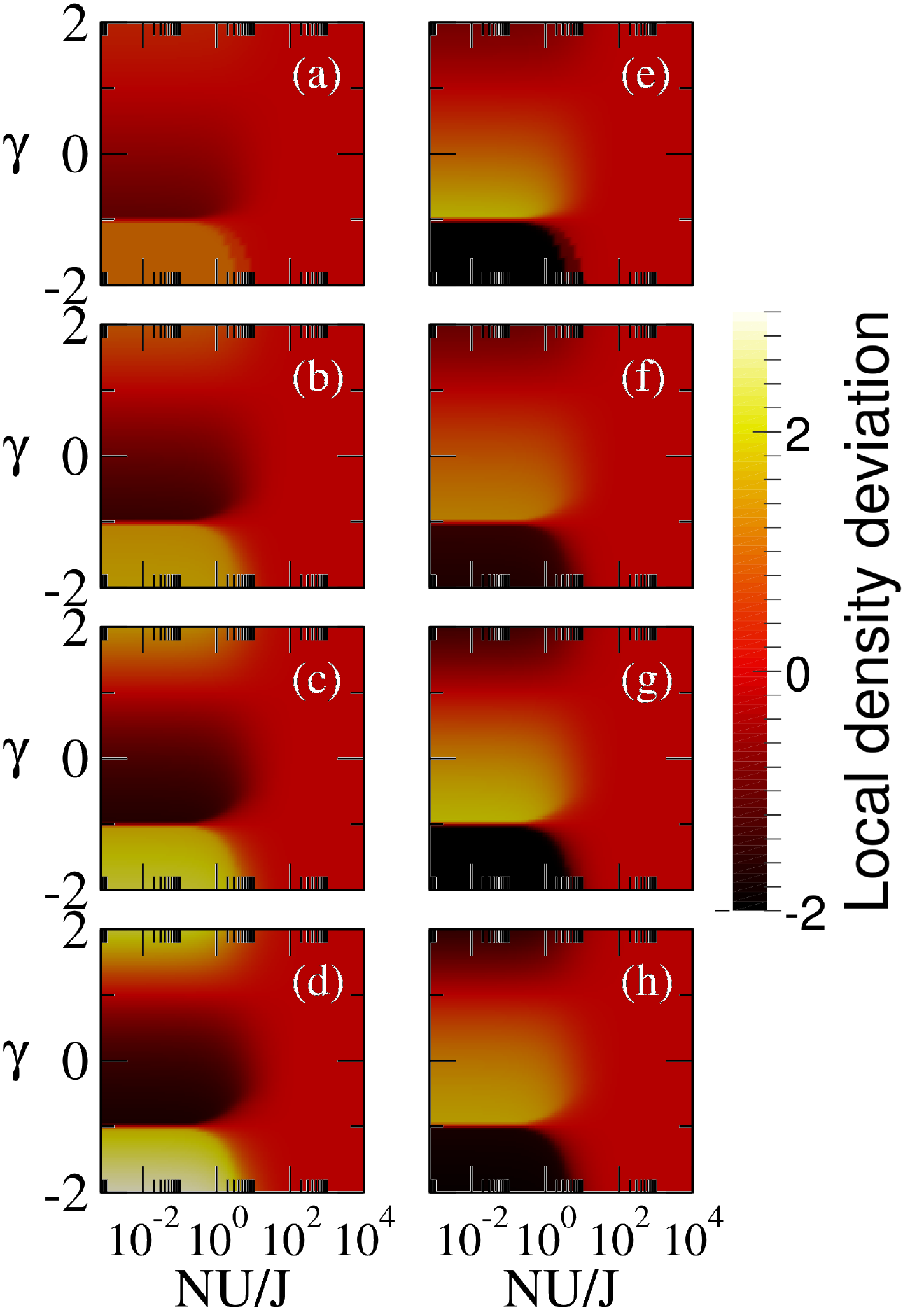}
\hspace{0.5cm}
\includegraphics[width=0.32\linewidth]{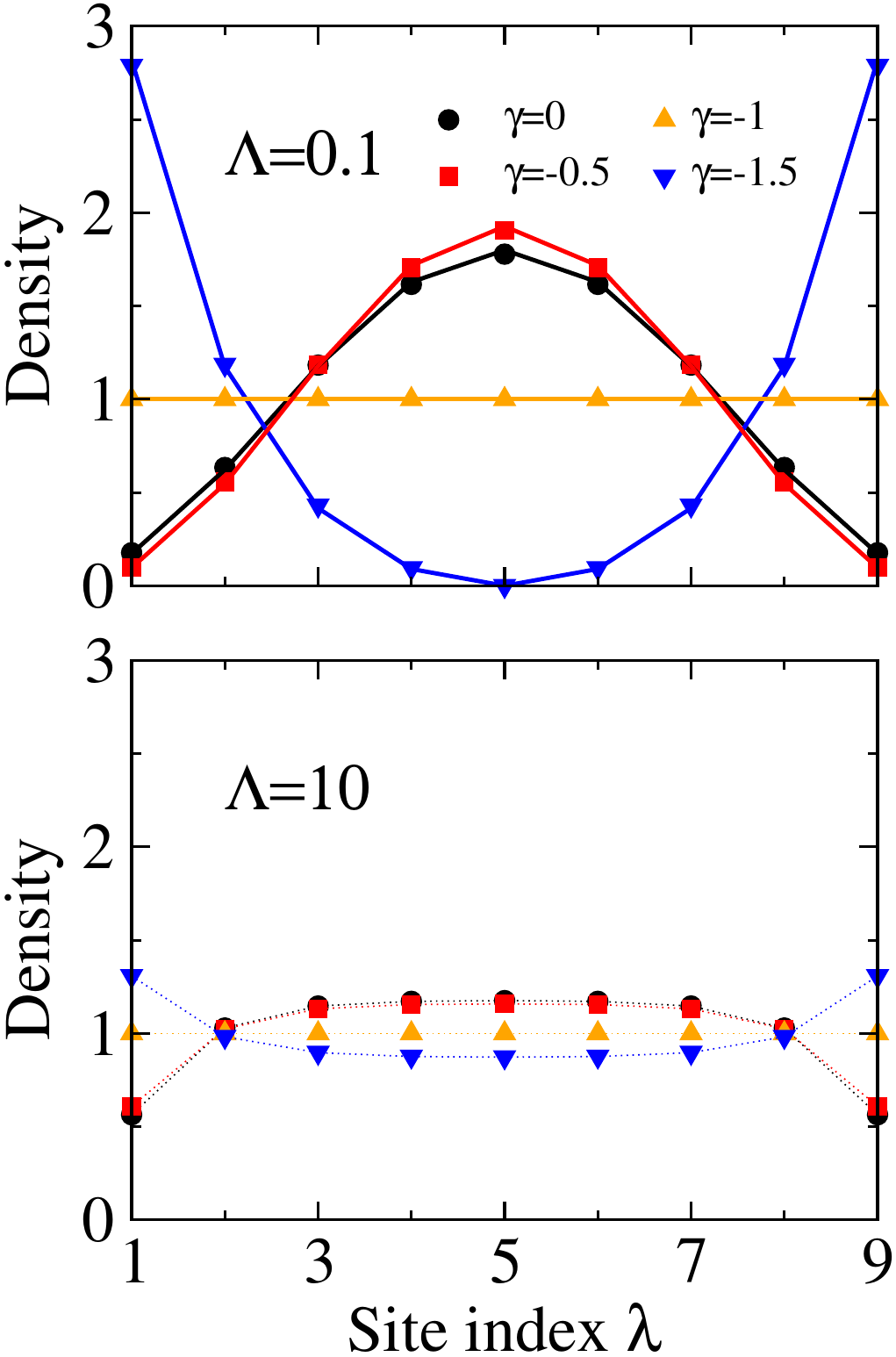}
\caption{(left panels) Deviation of the local density of the ground state 
from its global average at the tunable link (a-d), and in the center of the 
chain, furthest from the tunable link (e-h) as a function of 
$\Lambda\equiv NU/J$ and $\gamma$. The filling factor is the same for all 
panels, $N/M=2$. $M=3$, 4, 5 and 6, are depicted in panels (a,e), (b,f), 
(c,g) and (d,h), respectively. (right panels) Local average density of 
the ground state, $\langle \hat{a}^\dagger (\lambda) \hat{a}(\lambda)\rangle$, 
as a function of the site index $\la$ for $N=M=9$. Symbols depict the exact 
results for the values of $\gamma=-1.5$ (down-triangles), $-1$ (up-triangles), 
$-0.5$ (squares) and $0$ (circles). For the small interaction case, $\Lambda=0.1$, 
we add the non-interacting result, given explicitly in Eq.~(\ref{eq:den1}) for bulk 
states, as solid lines. The 
dotted lines for $\Lambda=10$ are just linear interpolations of the exact results.}
\label{den9}
\end{figure}

Similar expressions can be written when $\gamma < -1$, now using the 
odd single-particle solutions. The exact results shown in Fig.~\ref{den9} 
for $\Lambda = 0.1$ are almost identical to these simple predictions, 
except when $\gamma$ is very close to $-1$ where the breaking of 
the degeneracy of the odd and even solutions when $U \ne 0$ has 
to be taken into account.

The effect of varying $\gamma$ and $\Lambda$ on the density of the cloud is 
also particularly pronounced. For $\gamma=1$, the population is equal within 
all the sites, as the system is rotationally symmetric. For $\gamma=-1$ the 
situation is different due to the quasidegeneracy at the ground state level. For
small values of the interaction, the ground state is well represented 
by the cat states (\ref{cat}), which have an equal amount of fluxon and 
antifluxon components resulting also in a constant density along the chain, 
see Fig.~\ref{den9}. In the Mott regime, the system is equipopulated again, 
regardless of the value of $\gamma$, see the $\Lambda\gtrsim 10$ results in 
Fig.~\ref{den9}.  Already 
for $|\gamma|\gtrsim 1.5$, the 
density approaches the one built from the surface modes described in the first 
section, with population peaked on the sites around the tunable link.
In Fig.~\ref{den9} we provide a broader picture for filling factor $2$. 
Again for large interactions, both the central density and the density at 
the extremes approach the $N/M=2$ limit. Interestingly, the 
region of macroscopic superposition of fluxon-antifluxon states reflects in 
an almost equipopulation of all sites for all values of $\Lambda$. As 
discussed above, away from $\gamma=-1$ and for lower interactions, 
the cat-like state is less robust, resulting in a higher density in the 
center (extremes) as $\gamma$ increases (decreases).  
Monitoring the density of the chain can therefore be a good indicator 
of the macroscopic superposition states expected. For instance, the 
chain could be initially prepared at small but nonzero interactions 
and $\gamma=1$. Turning the tunable link from 
$\gamma=1$ to $\gamma=-1$, the density in the center will grow, reaching 
very large values for $\gamma\simeq -1$. At $\gamma=-1$ the chain should 
again be equipopulated. This transition from 
having almost zero population in the extremes to equipopulation would 
signal the regime of macroscopic superposition of fluxon-antifluxon states.

\begin{figure}[t!]
\centering
\includegraphics[width=0.7\columnwidth,angle=-0]{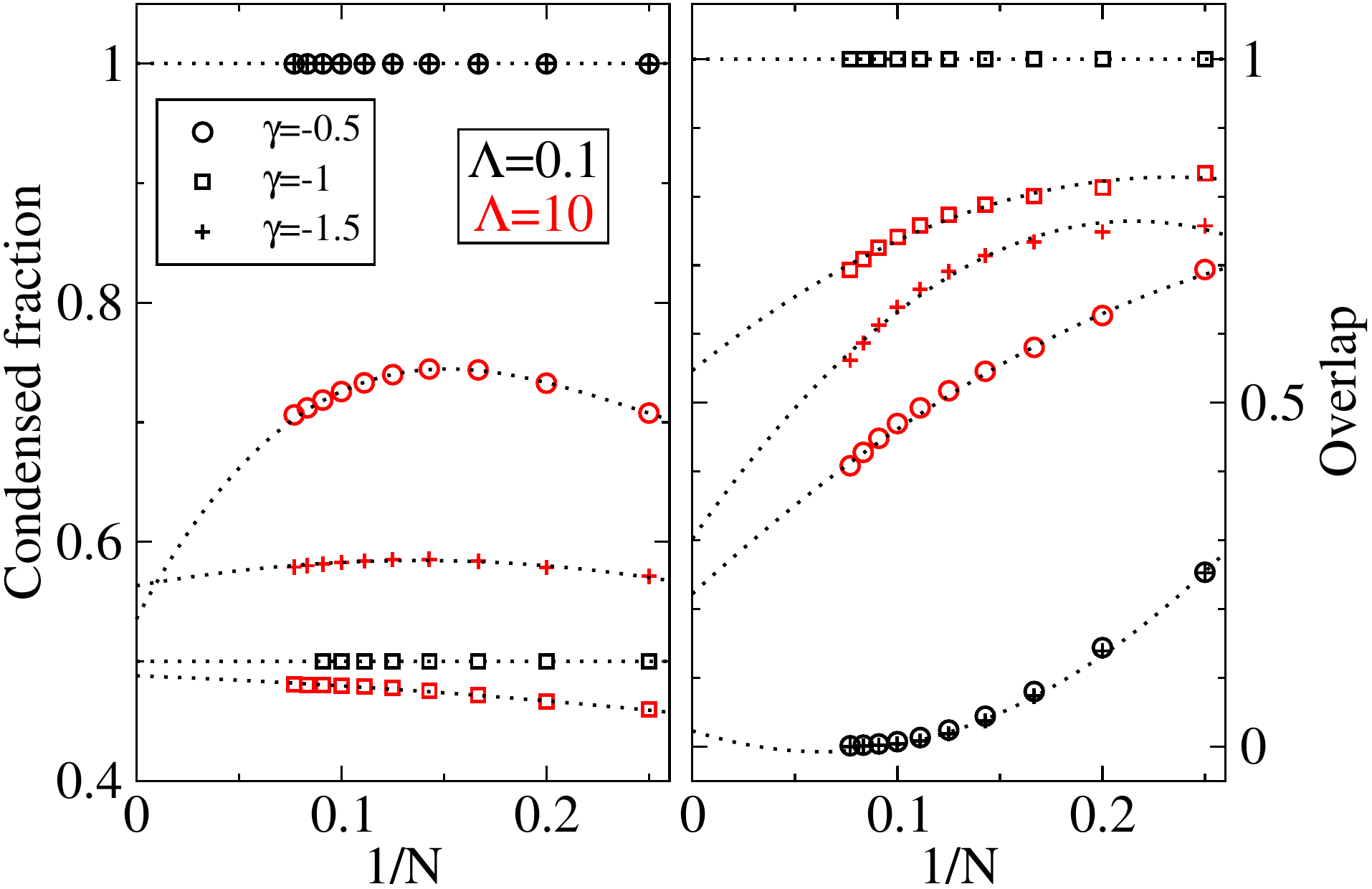}
\caption{Condensed fraction (left panel) and overlap determinant
(right panel), for $\gamma=-0.5$ (circles), $-1$ (squares) and $-1.5$ (crosses), 
with $\Lambda=0.1$ (black) and $\Lambda=10$ (red), as a function of $1/N$. 
The filling factor is one in all cases, and the maximum number of particles 
reached is $N=M=13$. The dotted lines depict a quadratic extrapolation of 
the calculated points to the thermodynamic limit 
($N\to\infty$).}
\label{vsn}
\end{figure}

The transition from a condensed to a fragmented state consisting in 
a macroscopic superposition of counterpropagating flows has been 
described for finite number of particles and sites using exact 
diagonalization techniques. It has been shown to take place at 
small but nonzero interactions, which actually helps to stabilize 
the superposition. Now we will stretch our numerical diagonalization 
techniques to explore the thermodynamic limit at small $\Lambda$. 
We have used the ARPACK package to go up to Hilbert spaces of 
dimension $\simeq 10^6$. This allows us to explore the behaviour of 
the macroscopic superpositions of semifluxon states with up to $N=M=13$, 
as shown in Fig.~\ref{vsn}. To roughly explore the thermodynamic limit we 
have performed extrapolations using up to quadratic terms in $1/N$,
shown in the figure with dotted lines. The transition to the macroscopic 
superposition phase is clearly seen to be much more robust at $\Lambda=10$ 
than $\Lambda=0.1$. In the 
former, for instance we find values of the overlap determinant of about 
$0.25$ for $\gamma=-0.5$, $-1$ and $-1.5$ in the 
extrapolation to the thermodynamic limit. 
The corresponding ground state fragmentation is close to 
$0.5$ in agreement with the predicted bifragmented nature of the 
superposition state. For smaller $\Lambda$ the 
superposition is only found at values $\gamma\simeq -1$: departures 
from it lead to a quick loss of the bifragmentation and to
small values of the overlap determinant.

\vspace{0.5cm}
\section{Summary and conclusions}
\label{sect:conclusions}

In this paper we have studied the role of a tunable link in a Bose-Hubbard 
chain of arbitrary size. In the case of non-interacting 
closed systems, we find counterpropagating persistent current states in the 
upper part of the spectrum with $\gamma=1$~\cite{Lee2006}, and in the ground state 
for the case of $\gamma=-1$, constituting a cat state of macroscopic flows. In 
this latter case, we have also analyzed the Bogoliubov excitations over 
mean-field states when interaction is present, by following a procedure similar 
to the one presented in Ref.~\cite{Paraoanu2003}. 

We have analyzed the robustness of these counterflow cat states by 
studying the energy spectrum, the density profile, the condensed fraction and the 
overlap of the non-interacting ground state manifold with the ground state computed  
by exact diagonalization, as a function of the interaction and $\gamma$. We have 
found that macroscopic superpositions of counterflow in weakly interacting Bose 
gases in a closed Bose-Hubbard chain with $\gamma=-1$ is more robust against 
fluctuation of the parameters, and is protected by a larger energy gap from 
excited states. 
Quadratic extrapolation to the large $N$ limit in the filling factor $N/M=1$ regime predicts 
the existence of such superpositions in the thermodynamic limit for small but 
non-zero interactions, $\Lambda\lesssim 5$.  

The production of macroscopic superpositions of semifluxon states in 
interacting many-body systems opens a new possibility to obtain persistent 
currents which may be useful in quantum computation and quantum simulation. An 
important feature of the results described in this article is that they 
are applicable to rings of any number of sites and thus can be engineered 
in a large variety of experimental setups. Thus, closed Bose-Hubbard chains 
with a single tunable link provide a versatile system in which macroscopic 
superpositions of flow states can be produced with a tunable twisted link, in 
which the tunnelling can be varied both in strength and phase. Among 
the possibilities that exist nowadays, cold atoms~\cite{Amico2014} and coupled 
non-linear optical resonators~\cite{Eichler2014} are promising quantum devices to produce 
countersuperflow states.

\section{Acknowledgments}

We acknowledge financial support from the Spanish MINECO (FIS2014-52285-C2-1-P and 
FIS2014-54672-P) and from Generalitat de Catalunya Grant 
No. 2014SGR401. A. G. is supported by Spanish MECD fellowship FPU13/02106.
B. J-D. is supported by the Ram\'on y Cajal MINECO program.

\vspace{2cm}

\clearpage
\bibliographystyle{iopart-num}
\bibliography{njp16bib}

 \clearpage
\appendix

\section{Many-body spectrum for weakly interacting systems: 
band structure when $\gamma = \pm 1$}
\label{Appendix-A}

In addition to the usual site basis for single bosons, it is convenient 
to consider also the flow basis that incorporates the matching condition, 
see Eq.~(\ref{eq:flow1}):
\begin{eqnarray}
|{\tilde \psi}_q\rangle 
&=&  {\tilde b}^{\dag}_q \,|{\rm vac}\rangle \nonumber \\ &=&  \frac{1}{\sqrt{M}} \sum_{l=0}^{M-1} e^{i \,l \phi_q} \,\hat{a}_l^\dagger \,|{\rm vac}\rangle \,,
\label{psi-q}
\end{eqnarray}
where $\phi_q=\frac{2 \pi}{M}q$ for $\gamma=1$, and 
$\phi_q=\frac{2 \pi}{M}(q+\frac{1}{2})$ for $\gamma=-1$,with $q= 0, 1, ..., M-1$. 
Since $e^{i \,l \phi_q}$ is a periodic function, $q$ is a cyclic index with period $M$. 

For example, for a trimer $M=3$ with $\gamma=1$, the flow basis, see Eq.~(\ref{eq:es5}), 
expressed in the Fock states (number of particles per each site), is:
\beqa
|0\rangle 
&=& {1\over\sqrt{3}} 
\left(|1,0,0\rangle + |0,1,0\rangle +  |0,0,1 \rangle\right) \\
|{\rm fl} \rangle 
&=& {1\over\sqrt{3}} 
\left(|1,0,0\rangle + e^{i {2\pi \over 3}} |0,1,0\rangle + e^{i {4\pi\over 3}} |0,0,1 \rangle\right) \nonumber\\ 
|{\rm afl} \rangle 
&=& {1\over\sqrt{3}}\left(|1,0,0\rangle + e^{-i {2\pi \over3}} |0,1,0\rangle + 
e^{-i {4\pi\over3}} |0,0,1 \rangle \right) \nonumber \,.
\eeqa
The basis elements correspond to equipartition of an atom in the three 
sites with zero flow, $|0\rangle $, a vortex state with clockwise flow $2 \pi$  
(fluxon), $|{\rm fl} \rangle $, and an anti-vortex (antifluxon) state, $|{\rm afl} \rangle$,  
with counterclockwise flow $-2 \pi$. Conversely, the $N$ boson Fock 
states in this basis are $|N_0,N_{\rm fl},N_{\rm afl}\rangle$, where $N_0,N_{{\rm fl}}$ and $N_{{\rm afl}}$
are the number of atoms in the $|0\rangle, |{\rm fl}\rangle$ and $|{\rm afl}\rangle$ flow 
states, respectively. It is important to stress that clockwise and 
counterclockwise states, $|{\rm fl}\rangle$ and $|{\rm afl}\rangle$ are degenerate 
since there is not a fixed rotation direction in the hamiltonian.

In the case of a trimer with $\gamma=-1$, the flow basis, Eq.~(\ref{eq:es5}), is:
\beqa
 |{\rm sf} \rangle 
&=&{1\over\sqrt{3}} 
\left(|1,0,0\rangle + e^{i {\pi \over 3}} |0,1,0\rangle + e^{i {2\pi\over 3}} |0,0,1 \rangle\right) \nonumber\\ 
|{\rm asf} \rangle 
&=& {1\over\sqrt{3}}\left(|1,0,0\rangle + e^{-i {\pi \over3}} |0,1,0\rangle + 
e^{-i {2\pi\over3}} |0,0,1 \rangle\right) \nonumber \\
|{\rm sfl+fl} \rangle 
&=& 
{1\over\sqrt{3}}\left(|1,0,0\rangle + e^{i \pi} |0,1,0\rangle + 
e^{i 2\pi} |0,0,1 \rangle\right)  \nonumber\\
&=&
{1\over\sqrt{3}} 
\left(|1,0,0\rangle - |0,1,0\rangle +  |0,0,1 \rangle\right) \,.
\eeqa
Here, the states $|\rm sf\rangle$ and $|{\rm asf}\rangle$ are degenerate and
correspond to a half-vortex (or semifluxon) and an anti-half-vortex (antisemifluxon) 
state with clockwise and counterclockwise flow $\pm \pi$, respectively. 
The third basis element $|{\rm sf+fl}\rangle$ is an excited state with clockwise 
flow which carries one and a half quantum flux of a vortex state: $3 \pi$. 

In this flow basis a general Fock state is given by $|N_{\rm  sf},N_{\rm asf},N_{\rm sf+fl}\rangle$, 
where $N_{\rm sf},N_{\rm asf}$ and $N_{\rm sf+fl}$ are the number of atoms in the 
$|{\rm sf}\rangle, |{\rm asf}\rangle$ and $|{\rm sf+fl}\rangle$ flow states, respectively.

In a Bose-Hubbard chain with $M=5$ sites and $\gamma=-1$, there are 5 elements in the 
flow basis: 
$|{\rm sf} \rangle \equiv |{\tilde \psi}_{0}\rangle$ and $|{\rm asf} \rangle \equiv |{\tilde \psi}_{-1}\rangle$ are 
degenerate and correspond to semifluxon and antisemifluxon states, respectively;
$|{\rm sf+fl} \rangle \equiv |{\tilde \psi}_{1}\rangle$ and $|{\rm asf+afl} \rangle \equiv|{\tilde \psi}_{3}\rangle$ 
that are degenerate and carry $\pm 3\pi$ flux; and $|{\rm sf+2fl} \rangle \equiv |{\tilde \psi}_{2}\rangle$ 
which is a non-degenerate state that carries $5\pi$ clockwise flow.

The band structure of the non-interacting spectrum and also for small 
interactions $\Lambda < 1$, see Figs.~\ref{spec} and ~\ref{fig-M3N4}, 
can be understood by means of the number of sites, the number of atoms, 
and the degeneracy of the flow basis elements.

For example for $M=3$ sites and $\gamma=-1$ the single-particle spectrum has 
a ground state with two degenerate eigenvectors, $|{\rm fl}\rangle$ and $|{\rm afl}\rangle$, 
and an excited state $|{\rm sf+fl}\rangle$. For a system with $N$ atoms, the 
total number of bands is $N+1$ that corresponds to the different possibilities 
of distributing $N$ atoms in a two level system with occupancies $N_{1/2}\equiv N_{\rm sf}+N_{\rm asf}$ 
and $N_{3/2} \equiv N_{\rm sf+fl}$, with the restriction $N=N_{1/2}+N_{3/2}$.
Inside each band, the number of quasi-degenerate levels can be calculated 
by counting the number of different possibilities $(N_{\rm sf},N_{\rm asf})$
to set $N_{1/2}$ atoms between two degenerate levels $|{\rm sf}\rangle$ 
and $|{\rm asf}\rangle$, where $N_{1/2} \equiv N_{\rm sf}+N_{\rm asf}$. 
The quasi-degeneration is $deg_{1/2}=N_{1/2}+1$.

In tables~\ref{table-3} and~\ref{table-4} we present the distribution 
in bands and the quasi-degeneracy inside each band, for the many-body 
spectrum of $N$  weakly-interacting atoms in $M$ sites.

Figure~\ref{fig-M3N4} shows the band structure of the
exact many-body spectrum for $N=4$ atoms in $M=3$ and $M=5$ 
sites, for small interactions. The band degeneracy predicted 
in Table~\ref{table-3} and~\ref{table-4} is in agreement with 
the quasi-degeneracy inside each band obtained numerically 
(see Fig.~\ref{fig-M3N4}).

\begin{table}[t]
 \begin{center}
      \begin{tabular}{| c || c | c | c | c | c |}
    \hline
    $M=3, N=4$ & band 1 & band 2 & band 3 & band 4 & band 5 \\
    \hline
    \hline
    $N_{1/2}=N_{\rm sf}+N_{\rm asf}$ & 4 & 3 & 2 & 1 & 0  \\
    \hline
    $(N_{\rm sf},N_{\rm asf})$ & {\bf (4,0)}&    (3,0)& (2,0) & (1,0)& (0,0) \\
                        & {\bf(0,4)}& (0,3)& (0,2) & (0,1)& (0,0) \\  
                       & (3,1)& (2,1) &  (1,1)      &            &       \\
                       & (1,3)& (1,2) &  (1,1)      &            &       \\         
                       &(2,2)&&&&\\
    \hline                   
    $deg_{1/2}=N_{1/2}+1$ &   5          & 4 & 3 & 2 & 1 \\         
    \hline
    \hline
    $N_{3/2}=N_{\rm sf+fl}$ & 0 & 1 & 2 & 3 & {\bf 4}  \\
        \hline
    $deg_{3/2}$           & 1 & 1 & 1 & 1 & 1  \\
    \hline
    \hline
    band degeneracy     & 5  &  4 & 3 & 2 & 1 \\
    $deg=deg_{1/2} \times deg_{3/2}$ & & & & & \\ 
    \hline
    \end{tabular}
    
    \caption{Distribution of $N=4$ atoms in the single-particle flow states of $M=3$ sites with $\gamma=-1$:
    $|{\rm sf}\rangle$, $|{\rm asf}\rangle$ and $|{\rm sf+fl}\rangle$. The many-body states in the flow basis
    are $|N_{\rm sf},N_{\rm asf},N_{\rm sf+fl}\rangle$.
    The mean-field like states are marked in boldface and correspond to: 
    $|4,0,0\rangle,|0,4,0\rangle,|0,0,4\rangle$.
    \label{table-3}}
 \end{center}
\end{table}

\begin{table*}[t]
 \begin{center}
\resizebox{\columnwidth}{!}{%
      \begin{tabular}{| c || c | c | c | c | c | c | c | c | c | c | c | c | c | c | c |}
    \hline
    $M=5, N=4$ & b1 & b2 & b3 & b4 & b5 & b6 & b7 & b8 & b9 & b10& b11 & b12 & b13 & b14 & b15 \\
    \hline
    \hline
    $N_{1/2}=N_{\rm sf}+N_{\rm asf}$ & 4 & 3 & 3 & 2 & 2 &1 &2&1&0&1&0&1&0&0&0  \\
    \hline
    $(N_{\rm sf},N_{\rm asf})$ & {\bf (4,0)}&(3,0)& (3,0)&(2,0) & (2,0) & (1,0)&(2,0)&(1,0)&(0,0)&(1,0)&(0,0)&(1,0)&(0,0)&(0,0)&(0,0)
                                 \\
                    & {\bf (0,4)}&(0,3)& (0,3)&(0,2) & (0,2) & (0,1)&(0,2)&(0,1)& &(0,1)& &(0,1)& & & \\
                    & (3,1)&(2,1)& (2,1)&(1,1) & (1,1) & &(1,1)&&&&&&&& \\
                    & (1,3)&(1,2)& (1,2)& &  & &&& && && & & \\
                    & (2,2)&& & &  & &&& && && & & \\
    \hline                   
   $deg_{1/2}=N_{1/2}+1$ &   5  & 4 & 4 & 3 & 3 & 2& 3&2&1&2&1&2&1&1&1 \\        
    \hline
    \hline
    $N_{3/2}=N_{\rm sf+fl}+N_{\rm asf+afl}$ & 0 & 1 & 0 & 2 & 1&3&0&2&4&1&3&0&2&1&0  \\
    \hline
     $(N_{\rm sf+fl},N_{\rm asf+afl})$    
     &(0,0)&(1,0)&(0,0)&(2,0)&(1,0)&(3,0)&(0,0)&(2,0)&{\bf (4,0)}&(1,0)&(3,0)&(0,0)&(2,0)&(1,0)&(0,0)\\
       &&(0,1)&&(0,2)&(0,1)&(0,3)&&(0,2)&{\bf (0,4)}&(0,1)&(0,3)&&(0,2)&(0,1)&\\               
       &&&&(1,1)&&(2,1)&&(1,1)&(3,1)&&(2,1)&&(1,1)&&\\          
       &&&&&&(1,2)&&&(1,3)&&(1,2)&&&&\\   
          &&&&&&&&&(2,2)&&&&&&\\       
      \hline
    $deg_{3/2}=N_{3/2}+1$           & 1 & 2 & 1 & 3 & 2&4&1&3&5&2&4&1&3&2&1  \\
    \hline
    \hline
    $N_{5/2}=N_{\rm sf+2fl}$ &0 &0&1&0&1&0&2&1&0&2&1&3&2&3&{\bf 4}\\
    \hline
    \hline
    \hline
band degeneracy     & 5  &  8& 4 & 9 & 6 &8&3&6&5&4&4&2&3&2&1\\
  $deg=deg_{1/2} \times deg_{3/2}$ & & & & &&&&&&&&&&& \\ 
    \hline
    \end{tabular}
}
    \caption{Distribution of $N=4$ atoms in the single-particle flow states of $M=5$ sites with $\gamma=-1$:
    $|\rm sf\rangle$, $|\rm asf\rangle$, $|\rm sf+fl\rangle$, $|\rm asf+afl\rangle$, $|\rm sf+2fl\rangle$. 
    The many-body states in the flow basis
    are  $|N_{\rm sf},N_{\rm asf},N_{\rm sf+fl},N_{\rm asf+afl},N_{\rm sf+2fl}\rangle$.
    The corresponding mean-field like states are marked in boldface and correspond to: 
    $|4,0,0,0,0\rangle,|0,4,0,0,0\rangle,|0,0,4,0,0\rangle, |0,0,0,4,0\rangle$, and 
$|0,0,0,0,4\rangle$. This case corresponds to a three level system with occupancies 
$N_{1/2}$, $N_{3/2}$ and $N_{5/2}$ with $N=N_{1/2}+N_{3/2}+N_{5/2}$ and 15 bands. 
    \label{table-4}}
 \end{center}

\end{table*}

\section{Detailed Bogoliubov analysis of $M=3$ and $M=5$ sites}
\label{appb}

In this appendix we provide a detailed analysis of the BdG spectrum 
for $M=3$ and $M=5$ sites at low interactions.

\subsection{Case 1: M=3 sites}

For $N$ atoms and $M=3$ sites, there are 3 macroscopically occupied states expressed 
in the Fock basis $|N_{\rm sf},N_{\rm asf},N_{\rm sf+fl}\rangle$ (see \ref{Appendix-A}), namely:
 $|{\tilde \Psi}_{q=0}^{(N)} \rangle =|N,0,0\rangle, \,$ $|{\tilde \Psi}_{q=-1}^{(N)} \rangle =|0,N,0\rangle$ and
$|{\tilde \Psi}_{q=1}^{(N)} \rangle =  |0,0,N\rangle$. The two first ones are degenerate and belong 
to the lowest band, and the  latter is the unique state in the highest band. One can 
consider a small rotational bias to break the degeneracy, such as the many-body ground 
state of the system is either the macroscopically occupied semifluxon mode 
$|{\tilde \Psi}_{q=0}^{(N)} \rangle=|N,0,0\rangle$, or the anti-semifluxon mode 
$|{\tilde \Psi}_{q=-1}^{(N)} \rangle =|0,N,0\rangle$.
When $M=3$, for a fixed mean-field state $|{\tilde \Psi}_{q}^{(N)} \rangle$,
there are three possible values of the index $k=-1,0,1$, and therefore 
two elementary excitations $k= \pm 1$.
In the BdG framework~\cite{Paraoanu2003},
they can be understood as the building up of a fluxon (vortex-like excitation)  when $k=1$ 
or an antifluxon (antivortex-like excitation) when $k=-1$.

Let us consider the macroscopically occupied semifluxon mode 
$|{\tilde \Psi}_{q=0}^{(N)} \rangle =   |N,0,0\rangle$, an excitation with $k=-1$ will lead 
the system to the excited state $|N-1,1,0\rangle$ which 
corresponds to the promotion of one particle from $|{\rm sf} \rangle$ to 
$|{\rm sf + fl} \rangle$, whereas a $k=1$ excitation will lead to the 
final state $|N-1,0,1\rangle$.

In Fig.~\ref{fig-M3N4} the BdG excitations are shown for a system with $M=3$, $N=4$
and $U/J=0.1$. When the macroscopically occupied mode is 
the semifluxon mode $|{\tilde \Psi}_{q=0}^{(N)} \rangle = |4,0,0\rangle$,
a $k=-1$ excitation leads the system to the excited state $|3,1,0\rangle$ which is the
third one in the lowest band, see Table \ref{table-3} in \ref{Appendix-A}. 
The BdG excitation energy is $E_{k=-1}^{q=0}/J=0.0969$ 
which is in good agreement with the excitation energy
calculated by exact diagonalization $\Delta E^{q=0}_{k=-1}/J=0.0992$ (see Table \ref{table-1}).

A $k=1$ excitation relative to $|{\tilde \Psi}_{q=0}^{(N)} \rangle =   |4,0,0\rangle$, leads to
the many-body state $|3,0,1 \rangle$, whereas a $k=-1$ excitation leads to $|3,1,0 \rangle$.
Both final states are degenerate in the fourth band (see Fig.~\ref{fig-M3N4}), and therefore
the BdG prediction gives the same excitation energies $E_{k=1}^{q=0}/J=E_{k=-1}^{q=0}/J=3.0969$ 
in quantitative agreement with the 
exact excitation energy $\Delta E^{q=0}_{k=1}/J=3.0998$ (see Table \ref{table-1}).

\begin{table}[t]
\begin{center}
\begin{tabular}{| c | c || c | c | c | c | c |}
    \hline
  $q$ & $|{\tilde \Psi}_{q}^{N)} \rangle$ & $k$ &$ E_k^{+}/J$ & $E_k^{-}/J$& $\Delta E^q_{k}/J$& $|\Psi_{q,k}'\rangle$\\
    \hline
    \hline
   0  & $|4,0,0\rangle$ & -1 & 0.0969 & \bcancel{-3.0969}& 0.0992 & $|3,1,0\rangle$ \\
     && 1 & 3.0969& \bcancel{-0.0969} &3.0998& $|3,0,1\rangle$\\
      \hline
  -1 & $|0,4,0\rangle$ & -1 & \bcancel{-3.0969} & 0.0969& 0.0992 & $|1,3,0\rangle$ \\
     && 1 & \bcancel{-0.0969}&3.0969  &3.0998 & $|0,3,1\rangle$  \\   
     \hline
  1 & $|0,0,4\rangle$ &-1 & \bcancel{2.8983} & -2.8983 & -2.8977 & $|1,0,3\rangle$\\
    && 1 & \bcancel{2.8983} & -2.8983 & -2.8977 & $|0,1,3\rangle$ \\
    \hline
     \end{tabular}      
\caption{Excitation energies of a macroscopically occupied mode $|{\tilde \Psi}_{q}^{(N)} \rangle$:
$E_k^{(\pm)}$ obtained from the Bogoliubov framework at a given $q$, and $\Delta E^q_{k}$ 
obtained by exact diagonalization. The excitation energies from the ground state 
$(q=0$ and $q=1)$ must be positive, whereas from the highest excited mode $(q=1)$ they 
must be negative. $|{\tilde \Psi}_{q,k}'\rangle$ is the resulting excited many-body state 
expressed in the Fock basis $|N_{sf},N_{\rm asf},N_{\rm sf+fl}\rangle$. $M=3, N=4, U/J=0.1$
\label{table-1}}
\end{center}
\end{table}

In Table~\ref{table-1} we compare the excitation energies calculated within the BdG formalism and 
the ones obtained by exact diagonalization of the BH hamiltonian for $M=3$ and $N=4$
and $U/J=0.1$. 
As can be seen in the table, the agreement for weakly interacting systems is very good.

\subsection{Case 2: $M=5$ sites}

For $N$ atoms in a BH chain with $M=5$ sites the flow basis is 
$|N_{\rm sf},N_{\rm asf},N_{\rm sf+fl},N_{\rm asf+afl},N_{\rm sf+2fl}\rangle$. There are 5 macroscopically 
occupied states ($q=-1,0,1,2,3$), see Table \ref{table-4} for $N=4$ atoms.
Two of them are degenerate in the lowest band (b1): $|\Psi_{q=0}^0 \rangle=|N,0,0,0,0\rangle$, and
$|{\tilde \Psi}_{q=-1}^{(N)} \rangle=|0,N,0,0,0\rangle\,$; two are degenerate in a middle band 
(b9) $|{\tilde \Psi}_{q=1}^{(N)} \rangle=|0,0,N,0,0\rangle$, and $|{\tilde \Psi}_{q=3}^{(N)} \rangle=|0,0,0,N,0\rangle\,$; 
and the last one is the highest excited state in the last band (b15) $|{\tilde \Psi}_{q=2}^{(N)} \rangle=|0,0,0,0,N\rangle \,$.
For a mean-field state, i.e. for a fixed $q$, the BdG framework will provides 
5 possible excitation energies labeled by $k=-1,0,1,2,3$ (or an equivalent set of values 
due to the periodicity). 

They can be understood as the building up of: a vortex-like excitation 
(fluxon) with $k=1$, an antivortex-like excitation (antifluxon) with 
$k=-1$, a doubly quantized vortex-like excitation (two fluxons) or 
equivalently three antifluxons with $k=2 \, (-3)$, and three fluxons or 
equivalently two antifluxons with $k=3 \,(-2)$.

Again, one can consider a small rotational bias such as the many-body ground 
state of the system is the macroscopically occupied semifluxon state 
$|{\tilde \Psi}_{q=0}^{(N)} \rangle=|N,0,0,0,0\rangle$. With another bias the ground 
state of the system can be $|{\tilde \Psi}_{q=-1}^{(N)} \rangle=|0,N,0,0,0\rangle$, 
that corresponds to the macroscopically occupied antisemifluxon state.

We have calculated the BdG excitations related to the macroscopically 
occupied states for $N=4$ atoms and $U/J=0.1$. In Table~\ref{table-2} we compare 
the BdG spectrum with the excitation energies obtained from the BH 
Hamiltonian. We have discarded all solutions which do not fulfill the BdG 
normalization~\cite{Paraoanu2003}. From this comparison, it follows that the BdG framework 
provides a good description for low-lying excitation energies when the 
interactions are small.

\begin{table}[t]
 \begin{center}
      \begin{tabular}{| c | c || c | c | c | c | c |}
    \hline
  $q$ & $|{\tilde \Psi}_{q}^{(N)} \rangle$ & $k$ &$ E_k^{+}/J$ & $E_k^{-}/J$& $\Delta E^q_{k}/J$& $|{\tilde \Psi}_{q,k}'\rangle$\\    
    \hline
    \hline
  0 & $|4,0,0,0,0\rangle$ &-1& 0.0584 & \bcancel{-2.2945} & 0.0590& $|3,1,0,0,0\rangle$ \\
    && 1 & 2.2945 & \bcancel{-0.0585} & 2.2913 & $|3,0,1,0,0\rangle$\\
    && 2& 3.6774 & \bcancel{-2.2954} & 3.6797 & $|3,0,0,0,1\rangle$\\
    && -2 (3) & 2.2954&\bcancel{-3.6774} & 2.2968 & $|3,0,0,1,0\rangle$\\  
      \hline
  1 & $|0,0,4,0,0\rangle$& -1 & \cancel{-1.4469} & -2.1711& -2.1828 & $|1,0,3,0,0\rangle$  \\    
     && 1 & \cancel{2.1711} & 1.4469& 1.4424 & $|0,0,3,0,1\rangle$ \\
     && -3 (2)& \cancel{2.1745}  & 0.0617 & 0.0569 & $|0,0,3,1,0\rangle$ \\
     && -2 (3) & \cancel{-0.0617} & -2.1744& -2.1822  & $|0,1,3,0,0\rangle$ \\   
     \hline
  2 & $|0,0,0,0,4\rangle$& -1 & \cancel{1.3206} &-1.3206 & -1.3205 & $|0,0,1,0,3\rangle$ \\
    && -4 (1) &  \cancel{1.3206}& -1.3206 & -1.3205 & $|0,0,0,1,3\rangle$\\
    && -3 (2) & \cancel{3.5575}&  -3.5575& -3.5583 & $|0,1,0,0,3\rangle$\\
    && -2 (3) & \cancel{3.5575}&-3.5575 & -3.5583 & $|1,0,0,0,3\rangle$\\
    \hline
     \end{tabular}
\caption{Same as Table \ref{table-1} for  $M=5, N=4$ and $U/J=0.1$. In this 
case the many-body states $|{\tilde \Psi}_{q}^{(N)} \rangle$ and $|{\tilde \Psi}_{q,k}'\rangle$ 
are expressed in the Fock basis $|N_{\rm sf},N_{\rm asf},N_{\rm sf+fl},N_{\rm asf+afl},N_{\rm sf+2fl}\rangle$.
\label{table-2}}
\end{center}
\end{table}

\end{document}